\documentclass[fleqn,usenatbib]{mnras}

\usepackage{newtxtext,newtxmath}

\usepackage[T1]{fontenc}

\DeclareRobustCommand{\VAN}[3]{#2}
\let\VANthebibliography\thebibliography
\def\thebibliography{\DeclareRobustCommand{\VAN}[3]{##3}\VANthebibliography}

\usepackage{graphicx}	
\usepackage{amsmath}	

\title[New satellite galaxy candidates of M104]{New dwarf galaxy candidates in the sphere of influence of the Sombrero galaxy}

\author[E.~Crosby et al.]{Ethan Crosby$^{1}$\thanks{E-mail: Ethan.Crosby@anu.edu.au},
Helmut Jerjen$^{1}$,
Oliver M\"uller$^{2}$,
Marcel S. Pawlowski$^{3}$ 
Mario Mateo$^{4}$
and Federico Lelli$^{5}$
\\
$^{1}$Research School of Astronomy and Astrophysics, Australian National University, Canberra, ACT 2611, Australia\\
$^{2}$Institute of Physics, Laboratory of Astrophysics, Ecole Polytechnique F\'ed\'erale de Lausanne (EPFL), 1290 Sauverny, Switzerland\\
$^{3}$Leibniz-Institut f\"ur Astrophysik Potsdam, An der Sternwarte 16, D-14482 Potsdam, Germany\\
$^{4}$Department of Astronomy, University of Michigan, 1085 S. University Ave., Ann Arbor, MI 48109, USA \\
$^{5}$Arcetri Astrophysical Observatory, Istituto Nazionale di Astrofisica, Largo Fermi 5, I-50125 Florence, Italy
}
\date{Accepted XXX. Received YYY; in original form ZZZ}

\pubyear{2022}

\begin{document}
\label{firstpage}
\pagerange{\pageref{firstpage}--\pageref{lastpage}}
\maketitle

\begin{abstract}

We report the discovery of 40 new satellite dwarf galaxy candidates in the sphere of influence of the Sombrero galaxy (M104) the most luminous galaxy in the Local Volume. Using the Subaru Hyper Suprime-Cam, we surveyed 14.4 square degrees of its surroundings, extending to the virial radius. Visual inspection of the deep images and GALFIT modelling yielded a galaxy sample highly complete down to $M_{g}\sim-9$ ($L_{g}\sim3\times 10^{5}\,L_\odot$) and spanning magnitudes $-16.4 < M_g < -8$ and half-light radii 50\,pc $< r_e <$ 1600\,pc assuming the distance of M104. These 40 new, out of which 27 are group members with high confidence, double the number of potential satellites of M104 within the virial radius, placing it among the richest hosts in the Local Volume. Using a Principle Component Analysis (PCA), we find that the entire sample of candidates consistent with an almost circular on-sky distribution, more circular than any comparable environment found in the Illustris TNG100-1 simulation. However the distribution of the high probability sample is more oblate and consistent with the simulation. The cumulative satellite luminosity function is broadly consistent with analogues from the simulation, albeit it contains no bright satellite with $M_{g}<-16.4$ ($L_{g}\sim3 \times 10^{8}\,L_\odot$), a $2.3\,\sigma$ occurrence. Follow-up spectroscopy to confirm group membership will begin to demonstrate how these systems can act as probes of the structure and formation history of the halo of M104.

\end{abstract}



\section{Introduction}

\subsection{Background}
The Lambda Cold Dark Matter ($\Lambda$CDM) model is the predominant and generally accepted cosmological model predicting the formation of galactic structures and has been well-studied through observations \citep{Planck2016,Abbott_et_al_2018,Planck2020,Alam2021,Brout_2022} and simulations \citep{Springel2008, Vogelsberger2014,EAGLE_2015,Griffen2016,TNG_MAIN}. Central to the $\Lambda$CDM paradigm is the mechanism through which collisionless dark matter forms gravitationally bound halos with a continuous spectrum of masses. Baryonic matter then generally becomes bound to these underlying dark matter halos in sufficient quantities to form observable stars and galaxies, from massive host $L_*$ galaxies to gravitationally bound satellite dwarf galaxies \citep{Moore1999,Klypin1999,Springel2008, GarrisonKimmel2014,Griffen2016,Kelley2019}. Large-scale cosmological simulations have provided a critical test-bed for detailed comparisons with the Universe as it is observed, providing the means to evaluate the importance of the various physical mechanisms that influence star-formation and the assembly of observable galaxies.

These comparisons have previously revealed numerous interesting and surprising discrepancies between observations and simulations. Here we focus on issues at scales involving single $L^*$ host galaxy and small galaxy group environments and the satellite dwarf galaxies which are bound to them, generally on the scales smaller than $1$\,Mpc. The most discussed issues historically included the ``Missing Satellites", ``Core-Cusp", and ``Too-Big-To-Fail" (\textit{TBTF}) problems (see \cite{Bullock2017} for a review), but have since been alleviated by the addition of baryon physics in cosmological zoom-in simulations \citep{Ogiya_2011,Buck2018,GarrisonKimmel2019,Wheeler2019}.

The inclusion of baryon physics alone however does not resolve a fourth problem known as the ``Disk-of-Satellites" or ``Satellite Plane" phenomenon. A satellite plane is a co-orbiting, aligned flattened distribution of satellite galaxies which is known to exist in the Local Group \citep{LyndenBell1976,Kroupa2005,Metz2007_a,Pawlowski2012,Ibata_2013,Conn2013,Pawlowski_2019} and the Centaurus A/M83 Group \citep{Tully2015,Mueller2018,Mueller_2019b,Kanehisa_2023}. Indications of satellite planes also exist in the NGC 253 \citep{MartnezDelgado2021}, M81 \citep{Chiboucas2013}, M101 \citep{Muller2017} and M83 systems \citep{Muller_2018_b}. All of these host galaxies together reside in the Local Sheet, a large-scale planar structure of nearby galaxy groups which are collectively moving towards the Virgo cluster \citep{Tully2008}. While surveys for planar systems outside of the Local Sheet are ongoing (such as MATLAS, \citet{Heesters2021}), nearly all have been confined to be within this cosmological structure.

A number of landmark studies find that observations of anisotropic, flattened satellite systems clash with results from $\Lambda$CDM simulations which do not commonly produce planar alignments of satellite galaxies about hosts, let alone display co-orbiting behaviour \citep{Metz_2009_a,Ibata_2013,Mueller2018,Pawlowski_2019,Pawlowski_2021_a}. Solutions have been put forward, firstly of a physical nature in which satellite galaxies are formed or accreted within the $\Lambda$CDM paradigm that are rare or not properly reproduced. Such proposals include galaxy accretion along flattened or narrow cosmic filaments \citep{Libeskind2010,Lovell2011}, group infall of dwarf galaxies \citep{Metz2009_B} and galactic fragments, or tidal dwarf galaxies, rotating in thin planes emerging from galactic interactions \citep{Kroupa_2005,Metz2007_b,Pawlowski2012,Kroupa_2012,Hammer_2013}. However these solutions present implications that are mostly unobserved, including the width of cosmic filaments and patterns in star formation histories (see \citet{Pawlowski2018} for a review). \citet{Cautun2015} made a number of statistical arguments, that the analyses of these extraordinary observations of planar satellite systems have been influenced by the \textit{"look-elsewhere"} effect and method selection bias. That is methods and results were tuned such that statistical significance was maximised and that when appropriately accounted for, the significance drops from $3-5\,\sigma$ to $\sim2\sigma$. More recent hypotheses found that the rare analogues in simulations that possessed a thin co-rotating satellite plane were transient and suggested the observed satellite planes are not stable structures \citep{Bahl_2014,Gillet_2015,Buck_2016,Shao_2019}. However, \citet{Pawlowski_2019} and \citet{Pawlowski_2021_b} found that proper motions from Gaia DR2 and HST data indicate that the co-rotating thin planes are stable. On the other side of the discussion, artificially elevated subhalo destruction in simulations for satellites with small orbital pericentres was used to suggest that the Milky Way satellite plane is a transitory structure consistent with $\Lambda$CDM simulations \citep{Sawala_2022}.

This discussion surrounding satellite planes is ongoing and has always been advanced by the introduction of more, and higher resolution proper motions and distances of satellite galaxies. However, a uniting and conclusive finding that explains satellite planes eludes the discussion, where often the presence or absence of a co-rotating plane relies solely on the motions of one or two satellites. More complete analyses that move beyond re-analysis of the same Local Group satellite galaxies is elusive, primarily due to data limitations, but the few larger scale comparisons suggest that the so far largely unaccounted for cosmological structures hosting nearby galaxies (namely; the Local Sheet) could introduce some bias \citep{Libeskind_2019}. How strongly does the cosmological web influence the structure of a satellite plane? Is the presence of a satellite plane dependant on host galaxy mass and morphology? Is the Local Group and the satellite planes it contains a cosmological outlier? These questions cannot be fully answered with our current understanding and data.

This is motivating further surveys of satellite galaxy systems, to work towards building a dataset free from statistical biases, and representative of the diversity of host galaxies, their satellite systems and the varied cosmological environments in which they reside.

\subsection{This Work}
In our study we search for satellite galaxies in the surroundings of the most massive galaxy in the Local Volume, the Sombrero galaxy (M104, also known as NGC4594 or PGC42407). We report 40 galaxies as new satellite candidates of M104 judged on their morphology, photometric properties, angular size and structural parameters.
We further derive stellar parameters for all 75 currently known satellite candidates using our deep Subaru Hyper Suprime-Cam imaging data. M104 is a luminous and massive galaxy of an unusual morphology with a dominant spheroid and a prominent disk (see Table\,2 of \cite{Kang_2022} for a compilation of morphological classifications), in many respects similar to Centaurus A. Historically, M104 has been regarded as an SA(s)a spiral galaxy \citep{RC3}, however recent analysis of its globular cluster system revealed a spatially segregated cluster population with a bimodal colour distribution more consistent with early-type galaxies, built through numerous accretion events with metal poor dwarf galaxies \citep{Kang_2022}. Additionally, M104 also possesses a huge stellar spheroid more consistent with elliptical galaxies rather than spiral galaxies \citep{Gadotti_2012}. It is suggested that M104 acquired its unique characteristic and iconic disk recently after a merger with a gas rich galaxy \citep{Diaz_2018,Kang_2022}.

M104 resides at a distance of $9.55\pm0.34$\,Mpc \citep{McQuinn_2016} from the Milky Way and is located in the foreground of the Virgo cluster southern extension \citep{Tully_1982,Kourkchi_2017}. It is largely isolated from other host galaxies and resides within the Local Volume, a spherical region of space with a radius of $\approx 10$\,Mpc around the Milky Way \citep{Karachentsev_2015_A}. \citet{Karachentsev2020} pointed out that "Many galaxies in the Virgo Southern Extension have radial velocities similar to that of Sombrero, but lie at greater distances typical of the Virgo cluster (15–20 Mpc)."

We estimate the virial radius $R_{200}$ of M104 using the equation:
\begin{equation}
	R_{200} = \sqrt[3]{\frac{3M_{200}}{4\pi\left(200\rho_{crit}(z)\right)}}
\end{equation}
from \citet{Kravtsov_2013} where $M_{200}$ is the mass within that radius and $\rho_{crit}(z)$ is the critical density of the universe as a function of redshift. $M_{200}$ is estimated using M104's stellar mass $M_{*}=17.9 \times 10^{10}$~M$_{\odot}$ \citep{Mu_oz_Mateos_2015} and the average $M_{*}/M_{200}$ ratio of 43.6 from selected galaxies in the \emph{TNG100-1} simulation \citep{Pillepich_2017,Nelson_2019}, where the selection criteria and simulation is described in more detail in Section \ref{M104_sat_plane}. This ratio from simulations agrees well with measured ratios from weak lensing, where \citet{Heymans_2006} measured $M_{*}/M_{vir,\,(z=0)}=34\pm12$. In Table \ref{tab:M104param} we use the $M_{*}/M_{200}$ ratio from the \emph{TNG100-1} simulation to estimate the virial mass.
We calculated the critical density at $z=0$ using the expression:
\begin{equation}
	\rho_{crit}(z=0) = \frac{3H_0^2}{8\pi\,G}
\end{equation}
To be consistent with large scale cosmological simulations, the adopted Hubble constant $H_0$ above is that presented in \citet{Planck2020} ($H_0=67.4$\,km\,s$^{-1}$\,Mpc$^{-1}$). With these parameters, the virial radius of M104 is estimated to be $R_{200}\approx 420$\,kpc. A complete list of basic parameters for M104 is given in Table \ref{tab:M104param}. 

\begin{table}
      \caption{Basic parameters of the host galaxy M104}
         \label{tab:M104param}
         \begin{tabular}{lll}
           \hline
           Morphology & SA(s)a & \cite{RC3} \\
           R.A.(J2000) & 12:39:59.4& \\
           DEC (J2000) & -11:37:23 &          \\
           v$_\odot$ & 1095\,km\,s$^{-1}$ & \cite{Tully_2016} \\
           $D_{25}$ & $8\farcm 7= 24.2\,$kpc & \cite{RC3} \\
           Distance (TRGB) & $9.55\pm0.34\,$Mpc & \cite{McQuinn_2016}*\\
           $(m-M)$ & $29.90\pm0.08$ & \cite{McQuinn_2016}\\
           $M_{B_T,0}$ & $-21.51$~mag & \cite{RC3}\\
           $v_{\rm rot}^{\rm max}$ & $345$~km\,s$^{-1}$ & \cite{Schweizer_1978} \\ 
           $M_{*}$ & $1.8 \times 10^{11}$~M$_{\odot}$  & \cite{Mu_oz_Mateos_2015} \\
           $M_{200}$ & $7.8 \times 10^{12}$~M$_{\odot}$ & this study\\
           $R_{200}$ & $420\,$kpc\ & this study \\
           \hline
         \end{tabular}\\
         (*): Distance measurements from various techniques are given in Table 2 of \cite{McQuinn_2016}.
\end{table}

\section{Observations}
We obtained CCD images of the M104 region in the \textit{HSC-g} band using the Hyper Suprime Camera \citep[HSC,][]{2018PASJ...70S...1M} at the 8.2m Subaru telescope at the Mauna Kea Observatories. The data acquisition was conducted as part of the observing proposal S18B0118QN (PI: H. Jerjen) on 2019 January 30-31. This proposal called for accompanying \textit{HSC-r2} imaging, however this could not be completed due to poor weather conditions during one night. The average seeing for the \textit{HSC-g} band observations was $1\farcs26\pm 0\farcs39$ arcsec. The HyperSuprime Camera is equipped with an array of 104 4k$\times$2k science CCD detectors, with an angular diameter of 1.5 degrees and a pixel scale of $0\farcs169$ at the centre of the field \citep{2018PASJ...70S...1M}. At the distance of M104 this angular diameter corresponds to a physical size of $\sim258\,$kpc.

\begin{figure*}
	\includegraphics[draft=false,width=16cm]{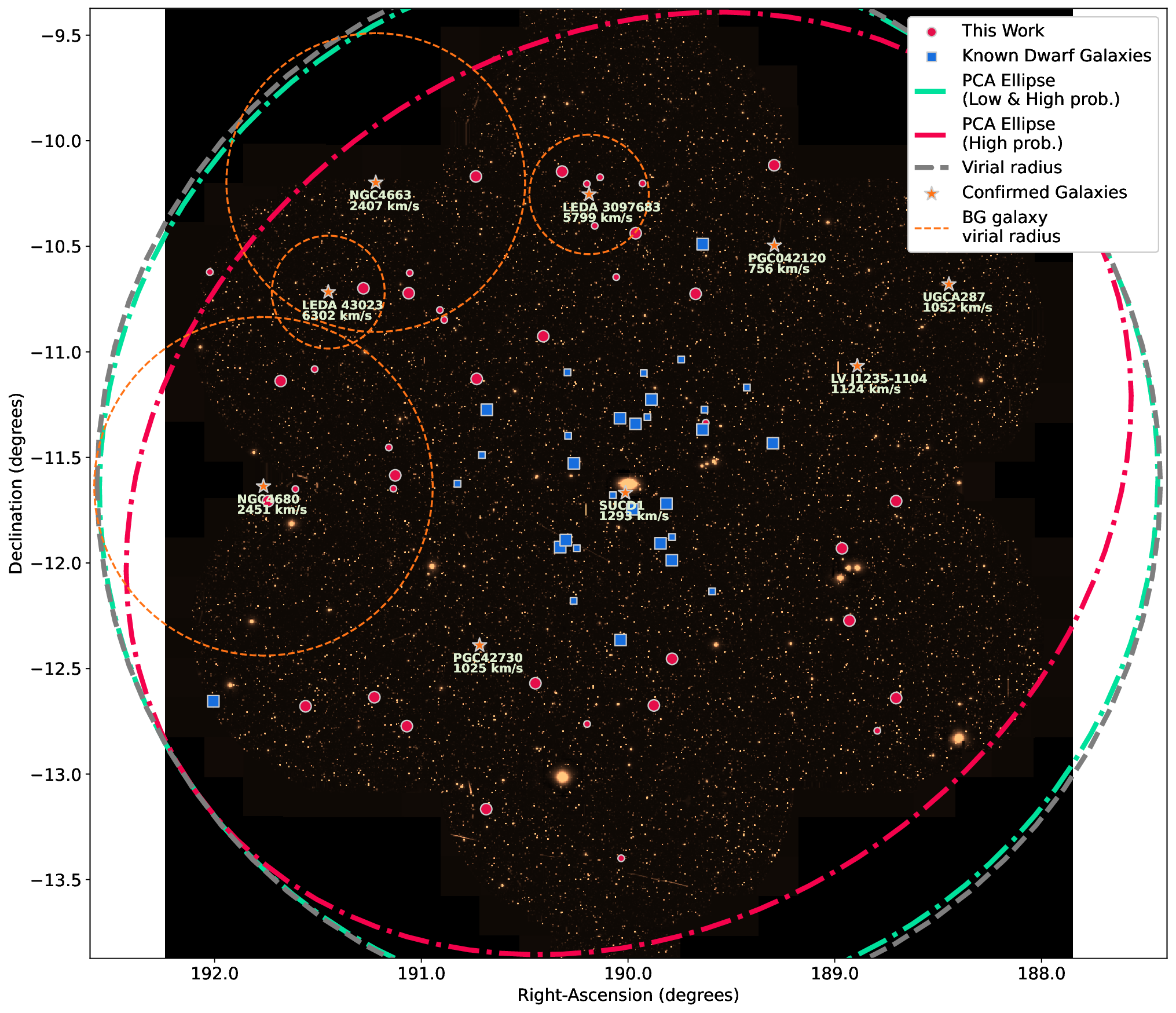}
	\caption{Map showing the HSC survey footprint centred around M104. Previously known dwarf satellite galaxy candidates are shown as blue squares, while new candidates from this work are the red circles. The size of the symbols is proportional to the M104 group membership probability as determined in our analysis (see Section \ref{visual_det}). Dark black areas are outside of the region captured by the telescope. The grey dashed line represents the approximate virial radius of M104, the green dashed-dotted line the best fit PCA ellipse of the 2D satellite distribution as described in Section \ref{M104_sat_plane}. The orange stars indicate the location of galaxies with measured LOS velocities and orange dotted circles denote the regions corresponding to the virial radius of background galaxies discussed in great detail in Section \ref{sec:m104_environment}, which may be the source of low probability candidates in that region.}
	\label{fig:M104_coverage}
\end{figure*}

During the observing run, seven HSC pointings were observed in a hexagonal pattern, with the central field on M104, extending the survey area out to a complete spherical volume with radius $\sim380\,$kpc at the distance of M104, which is approximately the virial radius of M104. Figure \ref{fig:M104_coverage} shows the survey footprint, nearby $L_*$ host galaxies, as well as known dwarf galaxies, and the new satellite candidates of M104. This observing configuration was chosen to achieve our scientific goal of detecting low surface brightness dwarf galaxies out to the virial radius of M104. Each HSC pointing consisted of three exposures: a short 30\,sec exposure for photometric calibration purposes and two 150\,sec exposures, dithered by half a CCD in R.A. and DEC directions.
This strategy allows us to detect satellite galaxies and low surface brightness dwarf galaxies down to a mean effective surface brightness of ${\sim\,27.5\,\text{mag}\,\text{arcsec}^{-2}}$ as shown in section \ref{sec:completeness}. 

The data reduction pipeline is described in greater detail in \cite{Crosby_2023}, but in short utilises the \textsc{hscpipe} software \citep{Bosch2018}, which is based on the pipeline being developed for the LSST Data Management system. The Pan-STARRS1 reference catalogue \citep{panstarrs} is used for photometric calibration, with uncertainty in the photometric zeropoint of $\sigma_{\text{ext}} = 0.087$\,mag.

\section{Known satellite galaxies and candidates}\label{known_satellites}
Previous studies have been conducted to search for satellite galaxies around M104 \citep{Karachentsev_2000,Javanmardi_2016,Carlsten_2020,Karachentsev2020}. \cite{Karachentsev_2000} started building a list of dwarf satellite candidates of M104 with analysis of the SERC EJ photographic plates, covering the entire field of M104 ($-18^\circ <\delta < 0^\circ$), but had a shallow surface brightness limit of $\mu \sim25-26\,\text{mag}\,\text{arcsec}^{-2}$. \cite{Javanmardi_2016} used a network of small diameter amateur telescopes to find satellite candidates in the central $0.7\deg\,\times\,0.7\deg$ ($\sim0.5\,{\deg}^2$) area around M104, with a surface brightness limit of $\mu \sim27.5\,\text{mag}\,\text{arcsec}^{-2}$. \cite{Carlsten_2020} used archival CFHT data, covering 70\% of the central 150\,kpc region ($\sim2\,{\deg}^2$) with a central surface brightness limit of $\mu_{i,0}\sim26\,\text{mag}\,\text{arcsec}^{-2}$. Our survey, while possessing a surface brightness limit of $\mu_{\text{lim}}=27.5$\,mag\,arcsec$^{-2}$ (see section \ref{sec:completeness} for details), has a total survey area of $\sim14.4\,{\deg}^2$.

However, these surveys have either small fields of view, or magnitude limits capable of detecting only the brightest dwarf galaxies. In total, there are 35 known candidates within our survey footprint. Photometric, structural and spatial parameters have been derived by us for all of them and are given in Table\,\ref{tab:photometry}. It is important to note that our HSC images extend to approximately one virial radius of M104, as we are searching for dwarf galaxies that are confidently bound to the halo of M104. Other surveys may be searching for dwarf galaxies in a wider volume, up to 2-3 virial radii from M104. In those cases it could be a survey of the M104 group, as in \cite{Karachentsev2020}, where some of those dwarf galaxies may in reality reside in the field as unbound field dwarfs, without being strictly bound to M104. In the case of M104, there also exists a nearby galaxy group with an approximate LOS velocity of $\sim1400$\,km\,s$^{-1}$, including NGC4700, NGC4742, NGC4781, NGC4804, and LEDA 43345, and may possess a wealthy system of satellites by itself. Extending the survey to the 2-3 virial radii regime, without LOS velocities for the dwarf galaxies in that region, risks misinterpreting dwarf galaxies as satellites of M104 which could actually be satellites of this background galaxy group.

Despite these challenges, \cite{Karachentsev2020} considered galaxies within an angular radial distance of $6^\circ$ or $\sim$1\,Mpc in the characterisation of M104's satellites galaxies, where within this region there may be $\sim10$ bright dwarf irregulars whose recessional velocities suggest they are satellites of M104 or at least members of the M104 group. For the purposes of this paper, we consider just the system of satellites within the 1 virial radius cutoff for consistency with our HSC images and to avoid the currently unmanageable challenges of associating dwarf galaxies to the correct host in the wider region.

Prior to our study there were five confirmed M104 satellites and 30 candidates within the virial radius limit. The confirmed dwarf satellites are: 
UGCA287 ($v_\odot=1052\pm 9$\,km\,s$^{-1}$ \cite{RC3}),
LV J1235-1104 ($v_\odot=1124\pm 45$\,km\,s$^{-1}$ \cite{Jones_2009}),
PGC042120  ($v_\odot=756\pm 2$\,km\,s$^{-1}$ \cite{Huchtmeier_2009}), 
SUCD1  ($v_\odot=1293\pm 10$\,km\,s$^{-1}$ \cite{Hau_2009}), and
PGC42730 ($v_\odot=1025\pm 45$\,km\,s$^{-1}$ \cite{Jones_2009}).
These 35 objects have been classified by us as five 'confirmed', 16 'high' probability and 14 'low' probability members. We describe what constitutes a 'high' or 'low' probability object in section \ref{visual_det}. The five unambiguous satellites of M104 were confirmed through follow-up spectroscopy. Of the remaining galaxies, still awaiting membership confirmation, the high probability candidates are predominately early-type dwarf galaxies, with a number of them possibly possessing bright nuclear star clusters (NSCs).

\section{Search for new M104 satellite galaxies}
In order to find new satellite galaxy candidates in the extended halo of M104, we search for unresolved, low surface brightness objects in our HSC observations. We employ this approach through a meticulous independent inspection of the entire survey area by three members of the team by-eye (as in \cite{Park_2017,Habas_2020,Muller_2020}). The entire data set is also reviewed multiple times and possible objects are logged and categorised based on their morphology. The images of known dwarf galaxies in the M104 system (Fig. \ref{fig:M104_candidates}) serve as a guidance to the appearance of potential new satellite galaxies. Our strategy is a conservative one, we only present satellite galaxies here that are unlikely to be false-positive candidates. With this strategy, we can expect a success rate of 60-80\% \citep{Mueller2018, Mueller_2019b}.

\subsection{Visual detection of dwarf galaxy candidates} \label{visual_det}

The full process for registering dwarf galaxy detections and the challenges associated with separating satellite candidates from background or foreground galaxies in the absence of distance and velocity measurements is described in our paper on the NGC2683 system \citep{Crosby_2023}, but to summarise here: satellite candidates are primarily detected through visual inspection of the morphology of extended objects resembling galaxies. Quenched early type dwarf galaxies such as dwarf Spheroidals (dSph) or dwarf Ellipticals (dE) are often located in high density galactic environments \citep{Binggeli_1987} or nearby a host galaxy and thus are often safely categorised as satellites. By the same phenomenon, star forming late type dwarf galaxies including dwarf Irregular (dIrr) and transition-type dwarf Transition (dTran) galaxies preferentially inhabit the low-density environments, also known as field, outside of the influence of a host galaxy, such that it is inherently more difficult to categorise these galaxies as satellites of a host galaxy as they could reside in the outskirts of a group \citep{Putman_2021}. Unresolved, small angular sized dwarf galaxies such as Blue Compact Dwarfs (BCD) or Ultra Compact Dwarf (UCD) are also likely to remain hidden as their morphologies closely resemble background galaxies or foreground stars. M104 is known to possess at least one UCD (SUCD1, $r_e=1.13$\,arcsec \cite{Hau_2009}), and one BCD (LV J1235-1104 \cite{Jones_2009}) companions. Any other similar satellite galaxies of M104 are likely to be missed in our survey. 
However, BCDs are thought to form only 5\% of the population of star-forming dwarf galaxies, being transient starburst systems \citep{Lee2009}, whereas UCDs are most probably the stripped nuclei of dEs \citep{Bekki2003}. Thus, even if our survey may be biased against BCDs and UCDs, we are not going to miss an important fraciton of the dwarf galaxy population.
Accounting for all of these factors, we apply a simple qualitative scheme to categorise the new detections between 'high' probability and 'low' probability of being satellites of M104. We use the following criteria to characterise a high probability satellite:
\begin{enumerate}
  \item The candidate lacks characteristic morphology of giant galaxies; spiral arms or \textit{cuspy} cores.
  \item The candidate exhibits expected surface-brightness vs. apparent magnitude ratios, as in Section \ref{sec:param_space}.
  \item The candidate half-light radius $>\,\sim6$ arcsec, or 300\,pc at M104.
  \item The candidate is visible in comparable alternative surveys, such as the DESI legacy survey \citep{Zou_2017,Zou_2018,DECALS}. This criterion helps to identify artefacts.
  \item The candidate possesses an extended low surface brightness component indicative of dwarf galaxies.
  \item The candidate is free from nearby foreground stars or background galaxies which could contaminate the image, or which the candidate may belong to (other than M104).
\end{enumerate}
A low probability object generally fails at least one of these conditions. Ultimately, membership probability is a qualitative measurement that is dependant on the author's experience, expectations and analysis methods and thus is generally subjective. To remove some of this subjectivity three members from our team conducted independent searches and each galaxy was carefully discussed afterwards before allocating a M104 group membership probability. In this paper, we have also adjusted our approach to categorising candidates to be consistent with previous reported discoveries of satellites around M104, by setting our selection criteria described above to be inclusive of confirmed satellites and candidates from prior papers, particularly such as those in \cite{Carlsten_2022}. For the purposes of making comparisons and generating mock catalogues, one can assume a high probability object has a 90\% chance and a low probability object a 50\% chance of being a true satellite galaxy of M104, based on the confirmation rates of follow-up observations from other surveys \citep{Chiboucas2013,Danieli_2017,Mueller2018,Mueller_2019b,Muller_2021}.

\subsection{Photometric modelling} \label{photometric_modelling}
We employed \textsc{galfit} 3.0.7 \citep{2010AAS...21522909P} to compute structural and photometric parameters for all candidate satellite galaxies. We use the S\'ersic profile \citep{Sersic1963} fitting functionality of \textsc{galfit} to model the light distribution of each object in the two-dimensional digital image. The analytical expression of the S\'ersic power law is:
\begin{equation}
	\Sigma(r) = \Sigma_{\text{eff}} \exp{\left[-\kappa \left(\left(\frac{r}{r_e}\right)^{\frac{1}{n}} -1 \right)\right]}
\end{equation}
where $r_{\text{e}}$ is the effective radius that contains half the total flux, $\Sigma_{\text{eff}}$ is the pixel surface brightness at the effective radius, $n$ is the S\'ersic index or concentration parameter, and $\kappa$ is a dependent parameter coupled to $n$.

For each model fit, \textsc{galfit} will produce two images: the best-fitting model and the model-subtracted residual. If the residual contains no evidence of the imaged galaxy and resembles closely the sky background, then the model is considered a good fit. In cases where galaxy light remains in the residuals, \textsc{galfit} allows us to fit multiple overlapping S\'ersic profiles to model the more complex morphology. For early type {dE/dSph} dwarf galaxies the best-fitting model typically consists of a single component S\'ersic profile, though an extra S\'ersic model may be required to account for a central nucleus, if present. Galaxies that have significant excess of stellar light present in the resulting residuals are generally transitional or irregular dwarf galaxies {dTran/dIrr}, where dust, star forming regions or tidal perturbations lead to asymmetric structures in their light distribution which cannot be modelled by concentric S\'ersic profiles. 
These galaxies require a combination of overlapping symmetric S\'ersic models to reproduce the observed complex light distribution. 
For galaxy light well modelled by a single S\'ersic model, the reported structural and photometric parameters in Table \ref{tab:photometry} are that of the single S\'ersic fit. Even for star forming {dTran/dIrr} galaxies, there generally exists an underlying extended and low-surface brightness component of the galaxy which hosts the asymmetric star forming regions. In cases with multiple S\'ersic profiles, the half-light radius, S\'ersic index and axis ratio are that of the underlying extended, low-surface brightness S\'ersic profile.

\begin{figure}
	\centering
	\includegraphics[draft=false,width=8.5cm]{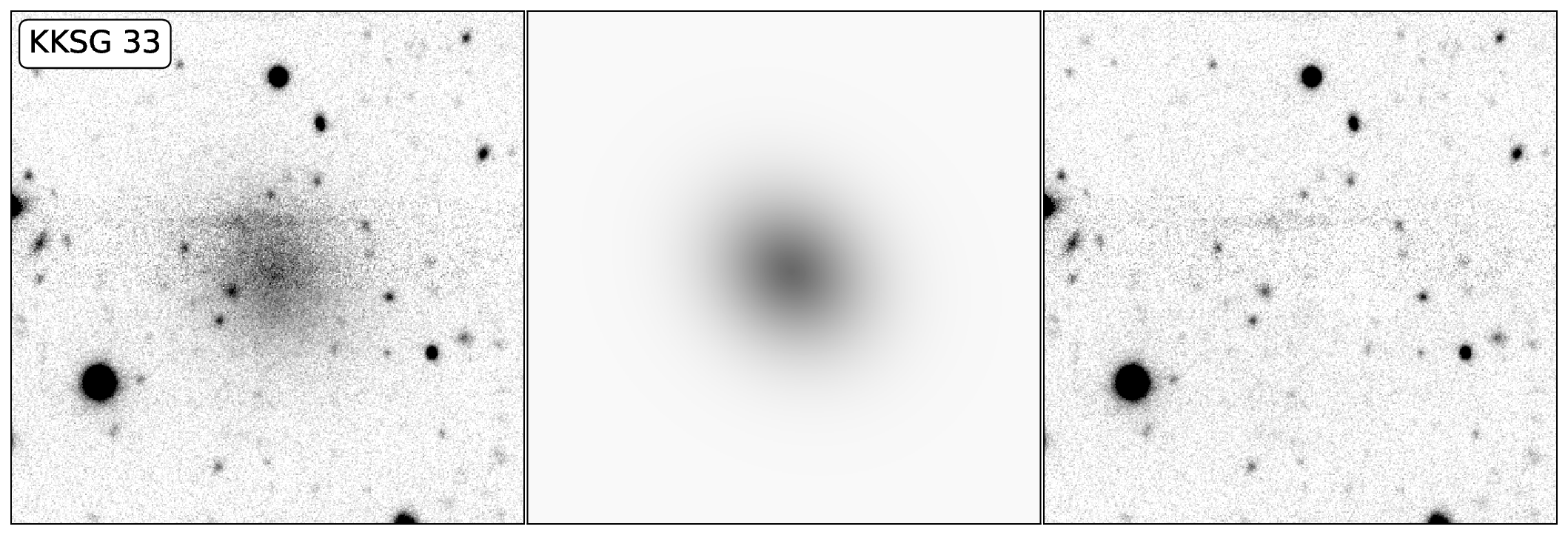}
	\includegraphics[draft=false,width=8.5cm]{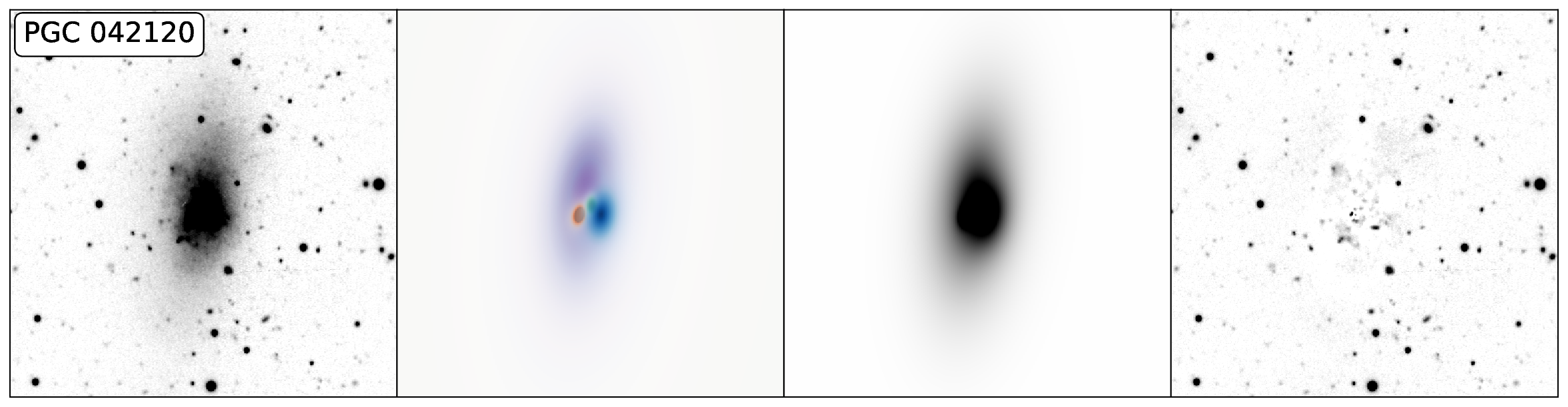}
	
\caption{Tri-frame images of a quenched dwarf spheroidal KKSG 33 (top row) and a star forming dwarf irregular PGC 042120 (bottom row). For KKSG 33, the left frame is the original HSC $g$-band image of the galaxy, the middle frame is the best-fitting \textsc{galfit} model and the third frame is the residual after the model is subtracted from the original image. For PGC 042120, we include an additional frame displaying the four individual constituent S\'ersic models in different colours, which summed together creates the total model shown in grey scale.}
\label{fig:sample_tri_frames}
\end{figure}

To demonstrate this process, we plot the tri-frame image consisting of the original image, the \textsc{galfit} model and the residual image for a single S\'ersic fit dSph, KKSG 33 and a multi S\'ersic fit dIrr, PGC 042120 in Figure \ref{fig:sample_tri_frames}. KKSG 33 is readily modeled with a single S\'ersic profile, while multiple overlapping S\'ersic profiles are required to fit the irregular star formation regions of PGC 042120 where we include an additional frame using colour to show the locations of four underlying S\'ersic models which create the full model for that galaxy.

\subsection{New Candidates}
\begin{figure*}
	\centering
	\includegraphics[draft=false,width=15.5cm]{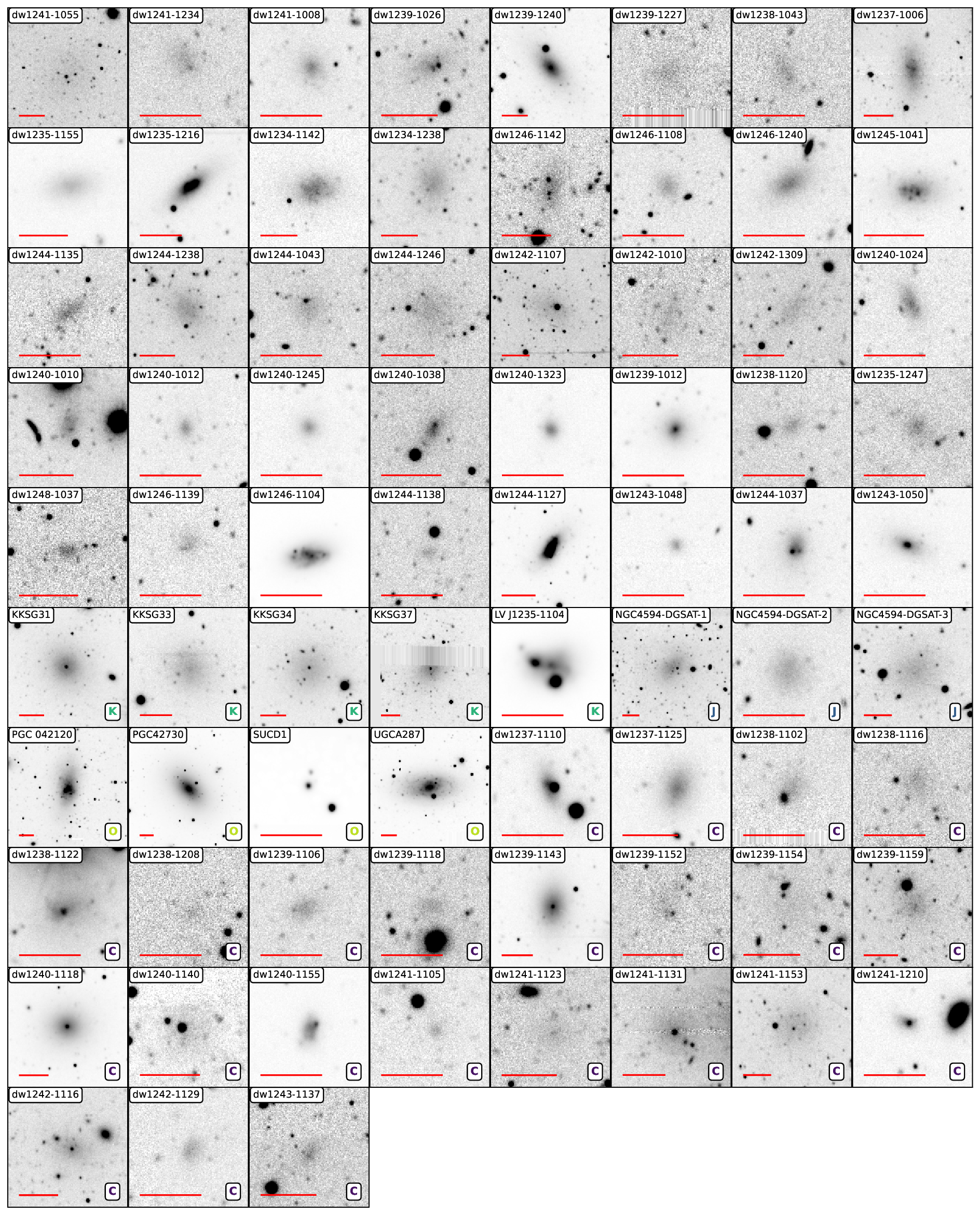}
\caption{HSC-$g$ band images of all 75 currently known M104 satellite galaxy candidates. North is up and East is left in each image. The red bar represents 1\,kpc at the distance of M104. A coloured letter in the bottom right corner indicates the initial paper in which this object was reported: (K): \citep{Karachentsev_2000,Karachentsev_2013}, (J): \citep{Javanmardi_2016}, (C): \citep{Carlsten_2022}, (O): other source. Images without a letter are our candidates.}
\label{fig:M104_candidates}
\end{figure*}

As a result of this process, we have found 40 new satellite galaxy candidates around M104, consisting of 23 'high' probability and 17 'low' probability candidates. We provide photometric and structural parameters with membership probabilities for these candidates in Table \ref{tab:photometry}. We assume each galaxy is at the distance of M104 to calculate the appropriate quantities. Follow-up observations involving velocity or distance measurements (using the Tip of the Red Giant Branch method from HST images for example) of the candidates is necessary to further justify this assumption. We display image cut-outs for these candidates in Fig.\,\ref{fig:M104_candidates}.

\subsection{Survey Completeness} \label{sec:completeness}
\begin{figure}
	\centering
	\includegraphics[draft=false,width=8cm]{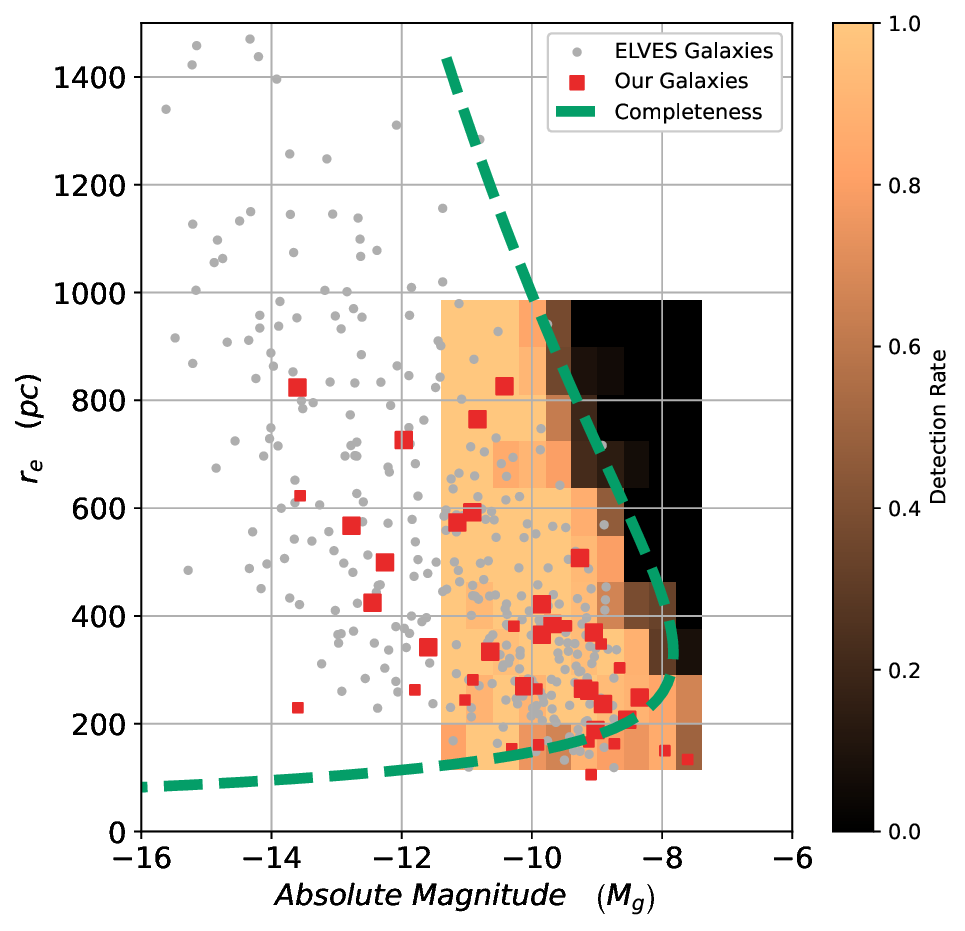}
	\caption{Survey completeness illustrated in the half-light radius-luminosity, $r_e - M_g$ plane. For comparison we plot galaxies from the Exploration of Local VolumE Satellites (ELVES) survey \citep{Carlsten_2022} with light grey points. The red squares are satellite candidates presented in this paper. The surface brightness limit is demonstrated with the detection rate from the underlying histogram, while the full analytic completeness is shown with the green line.}
	\label{fig:completeness}
\end{figure}

The completeness limits of our survey in terms of total magnitude, surface brightness and half-light radius are shown in Figure \ref{fig:completeness}. This plot consists of two parts; an underlying 2D-histogram and an analytic function. The histogram shows the galaxy detection rate as a function of angular size and total flux resulting from 2000 randomly generated dwarf spheroidal galaxies, consisting of a single S\'ersic profile, placed randomly throughout our HSC images of the M104 environment. A positive detection is registered if this simulated galaxy can be visually detected using the same manual process described in section \ref{visual_det}, verified by the independent analysis of two authors. This process includes reductions in detection rates due to obscuration of dim dwarf galaxies by bright background and foreground objects. These simulated galaxies consist of a single S\'ersic profile which has parameters in the range of $100\,$pc$\,\leq r_{\text{e}} \leq1000\,$pc, $-11.5\leq M_{\text{g}} \leq-7.5$ and are placed at the distance of M104. The axis ratio $b/a$ and S\'ersic index $n$ were allowed to vary from $0.5-1.0$. This histogram mirrors the surface brightness limit for our images, which is approximately $\sim27.5\,\text{mag}\,\text{arcsec}^{-2}$. Overlaid, we plot the analytic completeness relation from \cite{Ferguson_Sandage_1988}:

\begin{equation}
	m_{\text{tot}} = \mu_{\text{lim}} - \frac{r_{\text{lim}}}
	{0.5487 r_{e}}-2.5*\text{log}\left[2\pi \left(0.5958 r_{e}\right)^2\right]
\end{equation}

\noindent Where $\mu_{\text{lim}}$ is in $\text{mag}\,\text{arcsec}^{-2}$ and $r_{\text{lim}}$ in arcsec. 
The values best describing our observations are found to be $\mu_{\text{lim}}=27.5$\,mag\,arcsec$^{-2}$ and $r_{\text{lim}}=2$\,arcsec, which corresponds to the surface brightness limit discovered above, and the half-light radius cut-off. That is, any object smaller than this cut-off size cannot be discriminated from foreground stars or background galaxies.

For galaxies at the distance of M104 with half-light radii $r_{e}>300$\,pc, the analytic function fits the 50\% ridge line of the histogram well, but below this size limit it does not. This is a result of the manner in which the histogram was generated. It was based purely on whether the simulated galaxy is visible. However, it ignores additional considerations that take place to discriminate the object from background galaxies and foreground stars where at this size the candidate can be indistinguishable from those objects. Thus we consider the analytic relation to be a full representation of the real completeness limit, given both the visibility of the candidate or its distinction from other objects.

We therefore conclude that our M104 satellite galaxy survey is complete to a mean effective surface brightness of $\langle\mu{_e,g}\rangle \approx 27.5\,\text{mag}\,\text{arcsec}^{-2}$, which in total absolute magnitude, is 100 percent complete for objects more luminous than $M_{\text{g}}\approx -9$, and 50 percent complete for $-9<M_{\text{g}}<-8$, excluding compact galaxies with half-light radii smaller than $r_{\text{e}}=300\,$pc, which generally remain undetected.

\subsection{The M104 Environment} \label{sec:m104_environment}
In Section \ref{known_satellites} we described some of the challenges associated with identifying M104 satellites in the outskirts of M104, even though they have the hallmark of a dwarf galaxy. We can however be confident that dwarf galaxies found within one virial radius of M104 are satellites of M104 given this region is absent of nearby luminous host galaxies.

\begin{figure}
	\centering
	\includegraphics[draft=false,width=8cm]{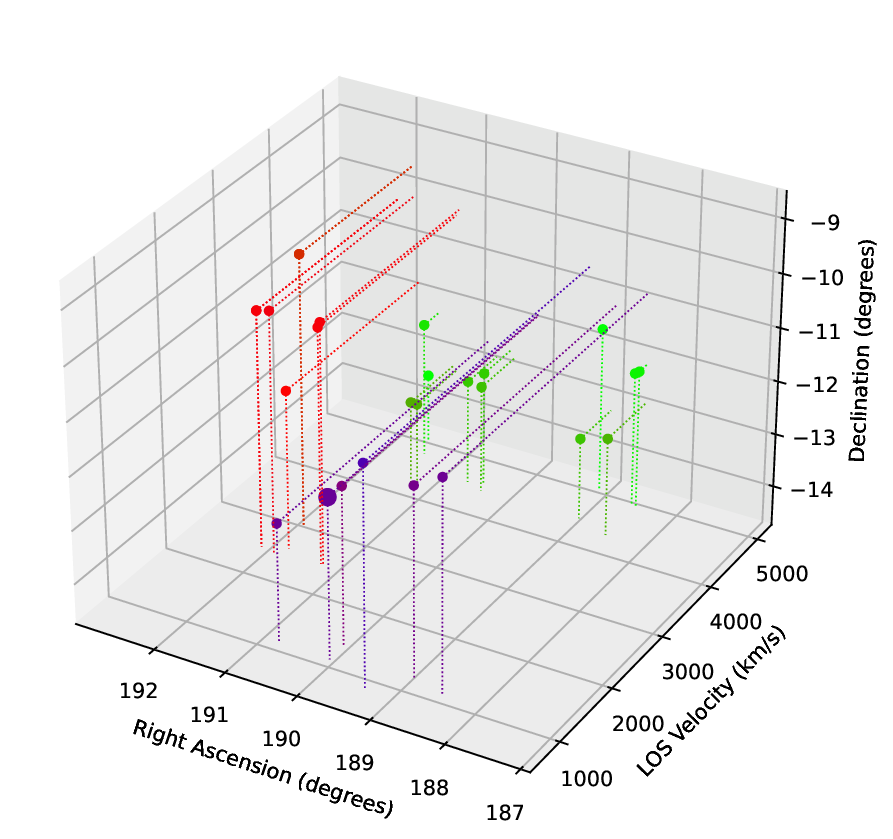}
	\caption{A three-dimensional scatter plot of all galaxies with known recessional velocities within our M104 survey footprint (x-z plane). M104 and its five satellites (purple coloured) with measured redshifts have a mean heliocentric velocity of 1056\,km\,s$^{-1}$ and a velocity dispersion of 175\,km\,s$^{-1}$. The larger of purple points is M104 itself. There is a significant velocity gap of over 1000\,km\,s$^{-1}$ to the next galaxy grouping in the background coloured in red, that includes NGC4663 (2407\,km\,s$^{-1}$) and NGC4680 (2451\,km\,s$^{-1}$). A more distant group in the background is coloured in green.}
	\label{fig:M104_vel_hist}
\end{figure}

Figure \ref{fig:M104_vel_hist} shows the three-dimensional scatter plot of all galaxies with known recessional velocities within our M104 survey footprint. M104 and its five satellites (purple coloured) with measured redshifts have a mean heliocentric velocity of 1056\,km\,s$^{-1}$ and a velocity dispersion of 175\,km\,s$^{-1}$. There is a significant velocity gap of over 1000\,km\,s$^{-1}$ to the next galaxy grouping at $~2500$\,km\,s$^{-1}$ coloured in red, that includes NGC4663 (2407\,km\,s$^{-1}$) and NGC4680 (2451\,km\,s$^{-1}$). Figure \ref{fig:M104_coverage} reveals that these two galaxies are at the north-eastern edge of survey footprint. Then a more distant group or cluster at $~4500$\,km\,s$^{-1}$, suggesting that each galaxy aggregate is separated by $15-30\,$Mpc from each other. Therefore, excluding isolated field dwarfs, any dwarfs belonging to one of these background groups should display significant differences in their angular size in comparison to the M104 satellites as they are much further away. However, we still identified a few low probability satellite candidates of M104 whose appearances make it difficult to definitively ascribe the candidate as belonging to a background group or to M104. 

In Figure \ref{fig:M104_coverage} we show the location and the estimated virial radius of several of these background galaxies which may be the true hosts of some of the low probability satellite candidates localised in that region of the survey area. Ultimately however, these background galaxies are sufficiently separated from M104 that we can, for the majority of dwarf galaxies, distinguish which host galaxy they belong to, with the exception of only a few, which we include as low probability candidates.

\section{Discussion}
\subsection{Candidate Parameter Spaces} 
\label{sec:param_space}
\begin{figure}
	\centering
	\includegraphics[draft=false,width=8cm]{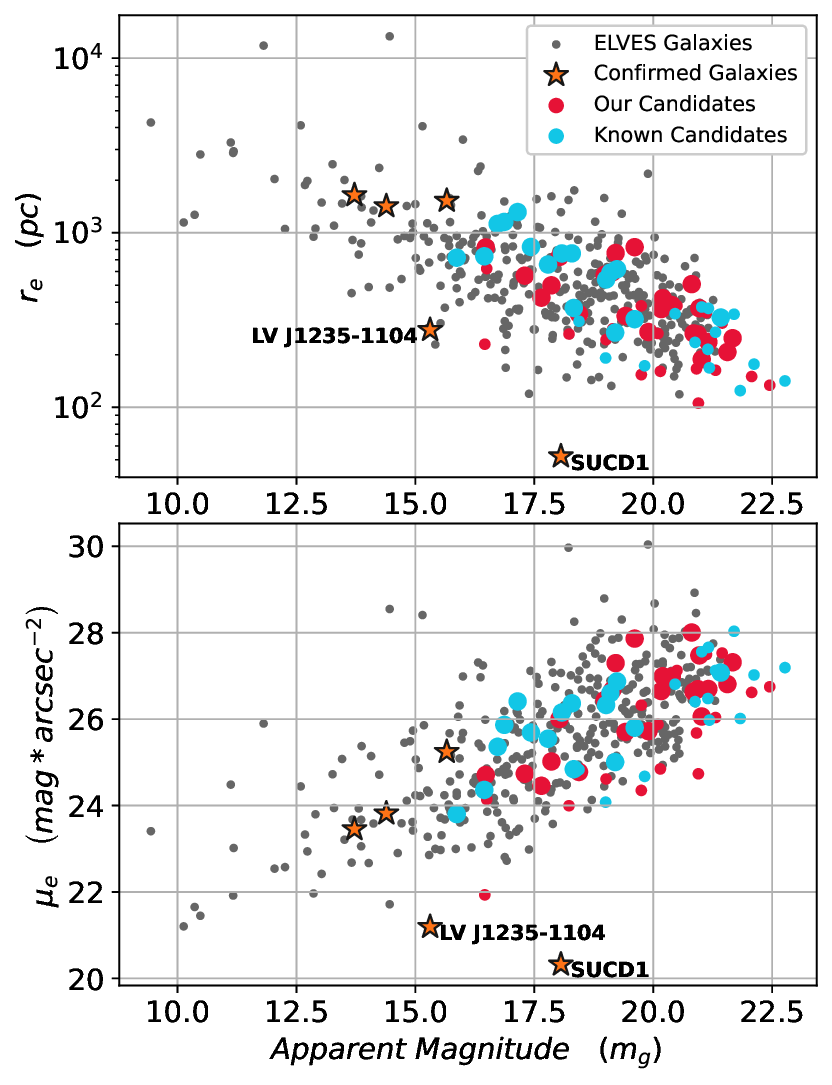}
	\caption{\textbf{Top panel:} The $r_e - m_g$ parameter space of galaxies in our sample and the ELVES dwarf galaxies. The half-light radius $r_e$ in pc is calculated assuming all candidates are at the distance of M104.
	\textbf{Bottom panel:} The $\mu_e - m_g$ parameter space of galaxies in our sample and the ELVES dwarf galaxies. 
	In both panels the red circles represent our satellite galaxy candidates and blue circles represent candidates and confirmed satellites already known. The size of the circles are scaled with membership probability (larger shape: high probability, smaller shape: low probability). The orange stars indicate the five satellites galaxies confirmed through spectroscopy. We also label the locations of the two confirmed compact satellite galaxies of M104, LV J1235-1104 and SUCD1, which do not follow the general trends in these plots.}
	\label{fig:parameter_space}
\end{figure}

We use the $\mu_e - m_g$ and $r_e - m_g$ parameter spaces as tools to assess the membership of the newly found galaxies in the M104 environment. Correlations exist in these three parameters for galaxies of all morphological classifications \citep{Kormendy1974, Caldern2020}. This correlation can be used to obtain qualitative reassurance about the candidacy of detected satellite galaxy candidates. This can help distinguish satellite candidates of M104 from galaxies far in the background ($d>50$\,Mpc), by highlighting unusually compact and high surface brightness galaxies which are more likely to be background galaxies. We compare these parameters of satellite galaxies in our sample with those in the Exploration of Local VolumE Satellites (ELVES) survey \citep{Zou_2017,Zou_2018,DECALS,Carlsten_2020,Carlsten_2021,Carlsten_2022} in Figure \ref{fig:parameter_space}, where we have assumed each candidate is at the distance of M104 in order to calculate $r_e$. From visual inspection of these plots, we find that our satellite candidates possess similar photometric and structural properties to other Local Volume satellites, which indicates that our candidates are sound choices. There are exceptions though, those being the compact galaxies SUCD1 ($v_\odot=1293\pm 10$\,km\,s$^{-1}$) and LV J1235-1104 ($v_\odot=1124\pm 45$\,km\,s$^{-1}$) as annotated on Fig. \ref{fig:parameter_space}, which are both spectroscopically confirmed satellites of M104. Such compact galaxies are, for reasons discussed in section \ref{visual_det} hard to distinguish from background galaxies, but also rare. Nonetheless may be neglecting the contributions of such compact galaxies to satellite systems, as has been suggested for the similar Centaurus A environment \citep{Voggel_2018,dumont2022}. Such galaxies can be very difficult to identify without additional information such as spectroscopy, or HST imaging, as ground based telescope images alone often lack the angular resolution to discriminate these galaxies from background galaxies or foreground stars.

\subsection{The Distribution of Satellites around M104}
\label{M104_sat_plane}
Satellite planes describe the observations of flattened disk like 3D distributions of satellites about their luminous host galaxies where the majority of the satellites co-move in those structures. Such configurations are observed in the Milky Way \citep{Pawlowski_2019}, M31 \citep{Ibata_2013}, Centaurus A \citep{Muller_2016_a,Muller_2018_b}, and indications of them exist in many other systems \citep{Heesters2021,Paudel_2021,MartnezDelgado2021}. Additionally, Gaia and HST proper motions of satellites in the Local Group continue to suggest satellite planes are stable co-rotating structures \citep{Pawlowski_2019,Pawlowski_2021_b}, which are scarcely represented in $\Lambda$CDM simulations. There still exists no broadly applicable and robust theories to explain the tension relating to all observed satellite planes.

To this end it would contribute to the ongoing investigation into satellite planes, particularly their frequency and sensitivity to their environment, to determine the spatial structure of satellites around the most luminous galaxy in the Local Volume, M104. To characterise a satellite plane, accurate 3D positions and line-of-sight velocities are required, the desired outcome of comprehensive follow-up observations based on the targets introduced in this study. Here we present 2D positions in the sky and begin to discuss the possible presence of a satellite plane.

To begin we measure the 'lopsidedness' in the distribution of the satellites, or the phase asymmetry of the satellites about the host. We define the lopsidedness as the length of the vector which is the average of the normalised vectors for each satellite. This way, it takes values between 0 and 1, where 0 represents a perfectly symmetric distribution of satellites, and 1 indicates the satellites all reside on a straight line originating from M104. There are numerous ways to quantify lopsidedness \citep{Pawlowski_2017,Gong_2019,Wang_2021,Samuels_2023}, but we choose this approach to minimise \textit{method-selection} bias, where the method is altered to extract maximum statistical significance and the \textit{look-elsewhere} effect. Both of these factors are persistent criticisms of the methods used to measure satellite planes \citep{Cautun2015}. These effects and biases may be present in alternative methods, like the opening angle method, where one can adjust the opening angle to whatever maximises statistical significance. Nonetheless, casual observations of the satellites in the satellite planes of M31 reveal what appears to be a lopsided distribution within that disk, so the presence of lopsidedness may indicate the presence of a satellite plane. Finally, as in other surveys limited to 2D sky coordinates \citep{Crosby_2023}, we use a Principal Component Analysis (PCA) to generate a best fit ellipse to the 2D coordinates of the satellites. The axis ratio of this ellipse is taken to be the flattening in the distribution of satellites.

\begin{figure}
	\centering
	\includegraphics[draft=false,width=8cm]{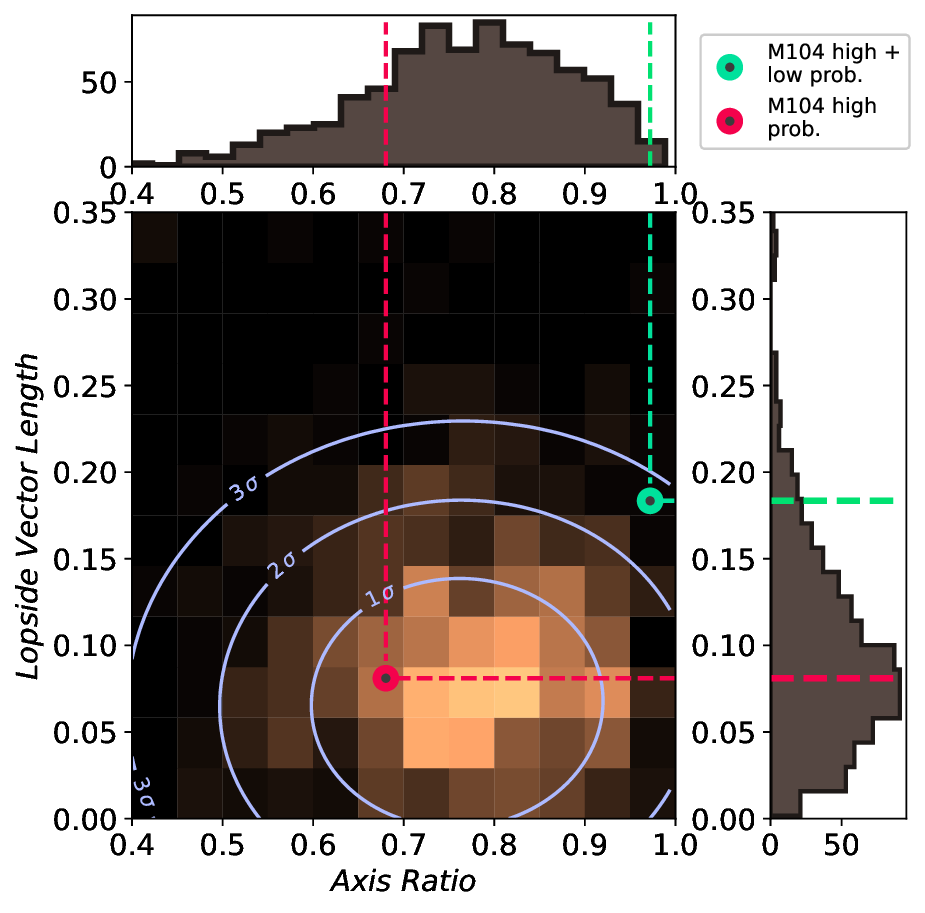}
	\caption{The 2D histogram formed from the PCA ellipse axis ratio and the Lopside Vector length, consisting of the 265 TNG100-1 simulated environments deemed similar to M104, as described in \ref{M104_sat_plane}. The green point is M104 with high and low probability satellite candidates, red point is M104 and high probability candidates only.}
	\label{fig:sim_axis_lop}
\end{figure}

The axis ratio of the PCA ellipse that includes all 75 satellite candidates of M104, independent of membership probability, is $b/a=0.97$, which is nearly circular and the lopsidedness is 0.18. Restricting the analysis to the high probability candidates only, the PCA axis ratio is $b/a=0.68$ and the lopsidedness is 0.08. To determine if these values are unusual or not, we compare the result to the Illustris TNG100-1 simulation \citep{Pillepich_2017,Nelson_2019}. The TNG100-1 simulation is a gravo-magnetohydrodynamical cosmological simulation with a side box length of 106.5\,Mpc, baryonic mass resolution of $1.4\times10^{6}M_{\odot}$ and dark matter mass resolution of $7.5\times10^{6}M_{\odot}$. This roughly corresponds to an absolute magnitude limit of somewhere between $M_g\sim-8$ and $M_g\sim-10$, which resembles our observational limits. From this simulation, we extract galaxy environments or FOF groups that are similar to the M104 environment at z=0 in terms of: 
\begin{enumerate}
    \item FOF group virial mass ($1-10\times10^{12}M_{\odot}$).
    \item sub-halo brightness, only constituent sub-halos with brightnesses comparable with our completeness limit are selected ($M_{g}<-9$ or $L_{g}\sim3*10^{5}\,L_\odot$).
    \item the number of satellites or sub-halos within the virial radius of the FOF group (30 to 100 satellites).
    \item isolation: the distance to the nearest FOF group ($>$2 times the virial radius).
\end{enumerate}
A total of 265 galaxy environments (FOF groups) in Illustris TNG100-1 match these conditions. We view each of these simulated environments three times along each Cartesian axis projecting the host galaxies onto a 2D plane and use the same PCA and lopsidedness analysis described above. We note that differences in the number of satellites between comparison samples can significantly bias results. Given we are uncertain of the true number of satellites around M104, we adopt to quoted range of 30-100 satellites in our sample to reflect the possible number in numbers M104 may truly possess. Future analyses must refine this range once candidates are confirmed with follow-up spectroscopy. As shown in Fig. \ref{fig:sim_axis_lop}, we find that while the lopsidedness of M104 is mostly consistent with simulations, the PCA axis ratio changes significantly depending on which satellites are included in the analysis. It varies from being nearly circular and a reasonable outlier from the TNG100-1 simulation if all 75 satellite candidates (high and low probability members), to being consistent with both if only the high probability satellites are considered.

We now investigate what might cause this extreme circularity. We compare M104 to simulated 'isotropic' systems of satellites. An 'isotropic' system of satellites consists of the same number of satellites around M104, but placed randomly in a 3D sphere and then projected to a 2D image. We perform this test $10^5$ times, measuring the PCA axis ratio and lopsidedness in this distribution for every iteration as above. The results of this experiment are shown in Fig. \ref{fig:sim_axis_lop}, which result in similar conclusions as to the TNG100-1 simulation, if all satellites are included, the M104 system is an outlier but if only the high probability satellites are included, it is consistent with simulations.

\begin{figure}
	\centering
	\includegraphics[draft=false,width=8cm]{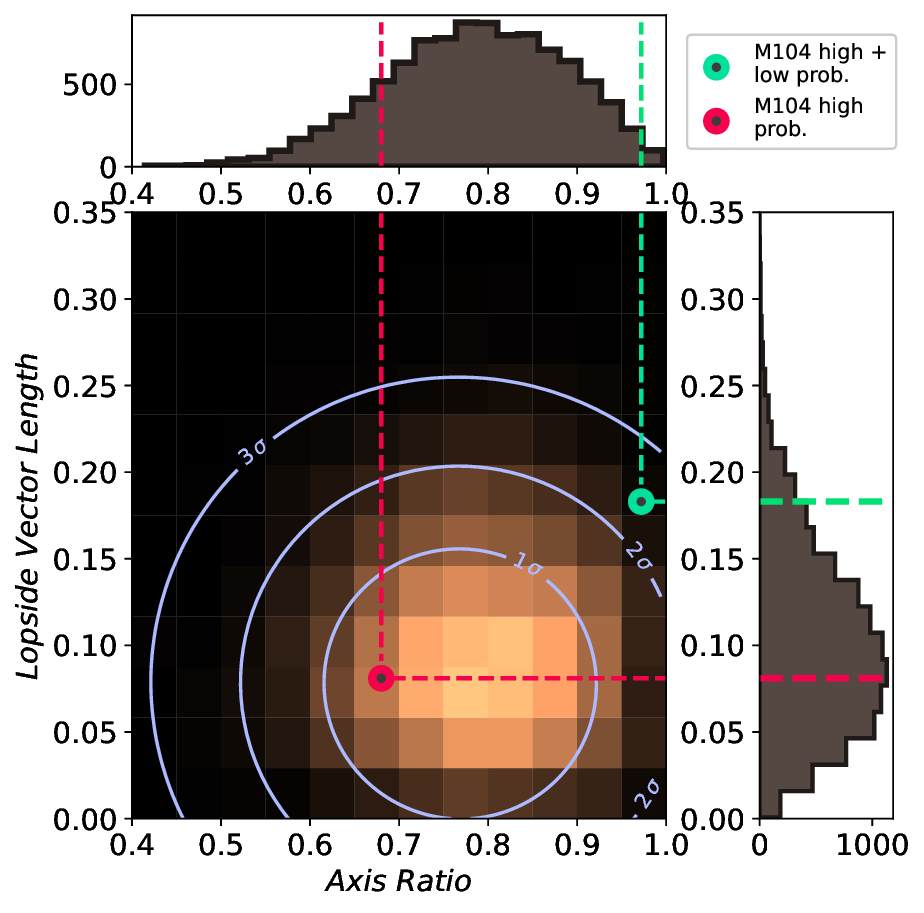}
	\caption{The 2D histogram formed from the PCA ellipse axis ratio and the Lopside Vector length, consisting of $10^5$ randomly generated isotropic systems similar to M104. The green point is M104 itself with all candidates, red is M104 high probability candidates only.}
	\label{fig:iso_axis_lop}
\end{figure}

Based on these results the interpretation is that the low probability satellite candidates may fill a fundamentally different distribution to the high probability satellites. The most likely explanation is that the subsample of low probability satellites includes dwarf galaxies that are bound to host galaxies like NGC4663 and NGC4680 in the background, a possibility we already mentioned over the discussion of the membership probability in the previous section. Removing all low probability members causes the 2D distribution of satellites around M104 to strongly resemble the distribution of similar systems in the Illustris TNG100-1 simulations. These different results emphasise the need for spectroscopy and resolved HST imaging follow-up observations of these candidates to unambiguously conclude if they are true satellites of M104 and fully understand the structure of the distribution of M104's satellites.

While we fail to find evidence of excess flattening in the 2D distribution of the satellites of M104 in comparison to simulations therefore inferring the presence of a satellite plane, further spectroscopy could still detect evidence of a satellite plane. In the most simple scenario, direct evidence of co-rotation could be detected if a plane is viewed edge-on. However, such a satellite plane could instead be viewed face-on. This scenario comes with a testable hypothesis; the line-of-sight velocity dispersion for this satellite system should be more tightly constrained than expected for a satellite system with random motions around a host galaxy of M104's mass, assuming the satellites are co-orbiting in that plane. Using our sample of M104-like systems from the TNG100-1 simulation, we measure a mean velocity dispersion of $\sigma=259\pm55\,km\,s^{-1}$ for subhalos within these systems. Measuring the velocity dispersion of the M104 satellite system and comparing against this predicted dispersion will be another interesting result from spectroscopic follow up of these candidates.

\subsection{Galaxy Luminosity Function of M104}

\begin{figure}
	\centering
	\includegraphics[draft=false,width=8cm]{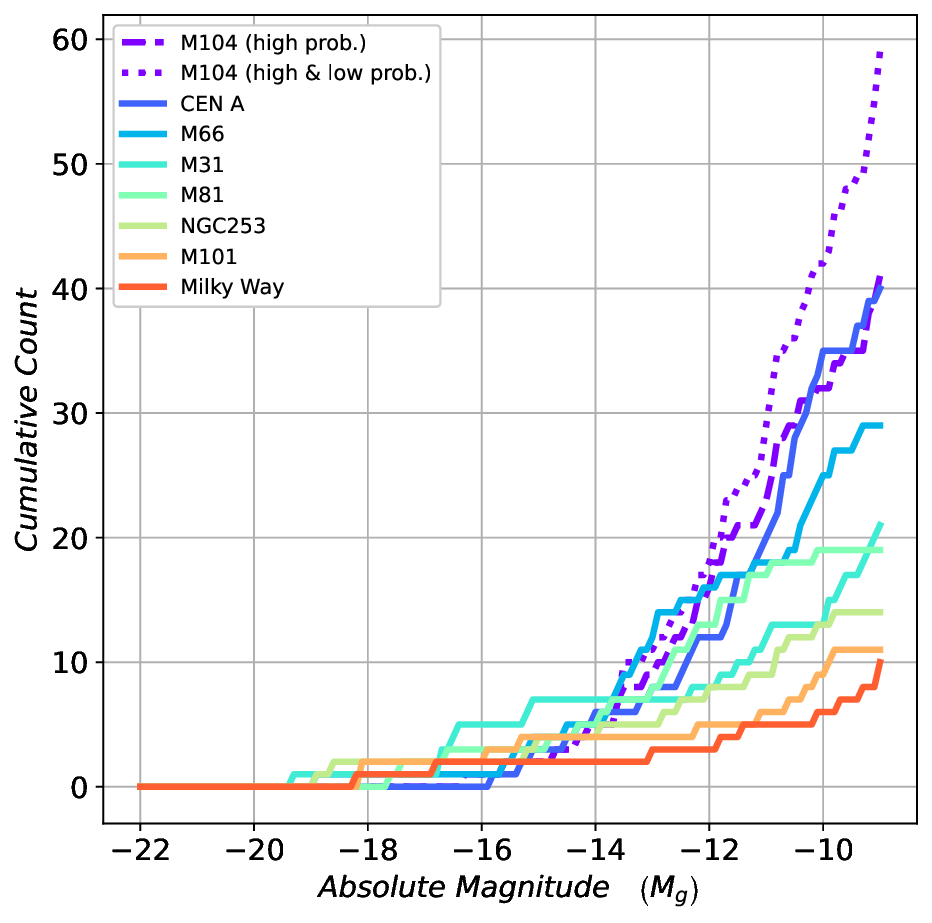}
	\caption{The cumulative satellite luminosity function (CSLF) of M104. The dotted purple line containing all candidates and the dashed line only high probability candidates. This is compared against seven Local Volume host galaxies as annotated in the figure.}
	\label{fig:GLF_LVcomp}
\end{figure}

We compare the cumulative satellite luminosity function (CSLF) of M104 to seven other nearby and well-known Local Volume galaxies for a qualitative comparison of the nature of the satellite galaxy systems and to compare the host $L_*$ galaxies. We show the CSLF in Figure \ref{fig:GLF_LVcomp}. For the purposes of this plot, we use data from both the ELVES catalogue \citep{Carlsten_2022} and the Catalog and Atlas of the LV Galaxies \citep{Karachentsev_2013} for the satellites and their magnitudes. For the ELVES catalogue we extract all known satellite candidates for M66 and M81. From the LV Atlas of galaxies we extract the candidates and confirmed galaxies within $400\,$kpc (as projected on the sky) for M101, NGC253, M31 and NGC253. For the MW, we include all confirmed satellite galaxies within a $400\,$kpc volume. Outside of the MW and M31, most of the reported satellites from these catalogues are unconfirmed candidates. All the CSLFs shown here could be considered upper limits on the CSLF for that host galaxy. Where $g$-band magnitudes are unavailable we use the SSDS band conversions \citep{Jester_2005} and approximate colours to generate $g$-band magnitudes. If we consider just the high probability satellite candidates of M104, we find the CSLF is very similar to Centaurus A (Cen A), which also bears a strong resemblance to M104 based on its morphology, both are elliptical/lenticular galaxies with a sluggishly star forming dust disk. Additionally, the brightest satellite of both Cen A and M104 has an absolute magnitude of $M_{g}\sim-16$ ($L_{g}\sim10^{8}\,L_\odot$), were the brightest satellite appears to be in the range of $-20\,<\,M_{g}\,<\,-17$ ($8.7 <\ $log$(L_{g,\odot}) < 10$) for the other, star forming spiral host galaxies.

\begin{figure}
	\centering
	\includegraphics[draft=false,width=8cm]{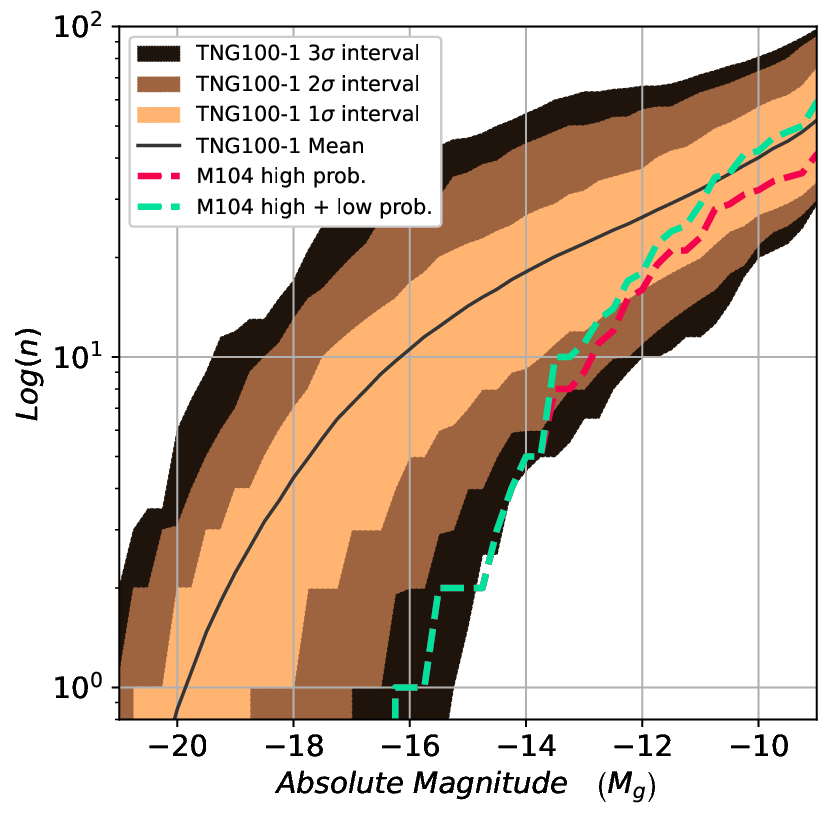}
	\caption{The cumulative satellite luminosity function (CSLF) of the high probability satellites of M104 (dashed red line) and all satellite candidates of M104 (green dashed line) compared against the mean CSLF for similar TNG100-1 simulated environments (solid black line) and the 1,2 and 3 $\sigma$ confidence intervals.}
	\label{fig:GLF_simcomp}
\end{figure}

We repeat this comparison with sample of 265 M104-analogue galaxy environments we described in \ref{M104_sat_plane}. In Figure \ref{fig:GLF_simcomp} we plot the M104 CSLF, and compare it against the CSLFs of the simulated M104-analogues. In the CSLF, M104 has a deficiency of satellites in the high luminosity end with magnitudes $M_g<-14$ ($L_{g}\sim10^{7}\,L_\odot$), though an expected amount of satellites with magnitudes $M_g>-14$ compared to the reference environments, even if only high probability candidates are included, indicating the result is robust even if all low probability candidates are discounted as M104 satellites. In this way, ignoring the small numbers of potentially missed candidates falling in CCD gaps or becoming obscured by bright objects, the high and low probability curve is an upper limit of the CSLF of M104 (for satellites within one virial radius). 98.7\% of the comparison environments possess a satellite galaxy within the virial radius brighter than $M_g=-16.4$ ($\sim3*10^{8}\,L_\odot$), which is the magnitude of M104's brightest satellite, therefore M104 is a 2.3$\,\sigma$ outlier in this regard. Note that this result may reflect the radial distribution of these high luminosity satellites, since we are observing satellites within the virial radii of M104 and simulations. In \cite{Crosby_2023} we reported an analogous anomaly, a significant magnitude gap of 6.1\,mag between the host galaxy NGC2683 and the brightest satellite KK69, in the NGC2683 system. We find a similarly large magnitude gap 5.1\,mag in the luminosity function of M104, despite the significantly more populated system of satellites which should increase the probability of finding high luminosity satellites.

\subsection{Radial Satellite Distribution of M104}

Recent research has indicated that the radial distribution of satellites about their host is not correctly reproduced in large-scale cosmological simulations, particularly for satellites which pass by close to their host in an orbit, otherwise described as a small pericentre \citep{Guo2013,vandenBosch2018,Webb2020}. In these simulations, the choice of fundamental parameters including the simulation resolution, the dark matter particle mass and the softening length leads to artificially elevated subhalo destruction, accretion and disruption. This manifests as a reduced density of subhalos at small pericentres compared to observations.

\begin{figure}
	\centering
	\includegraphics[draft=false,width=8cm]{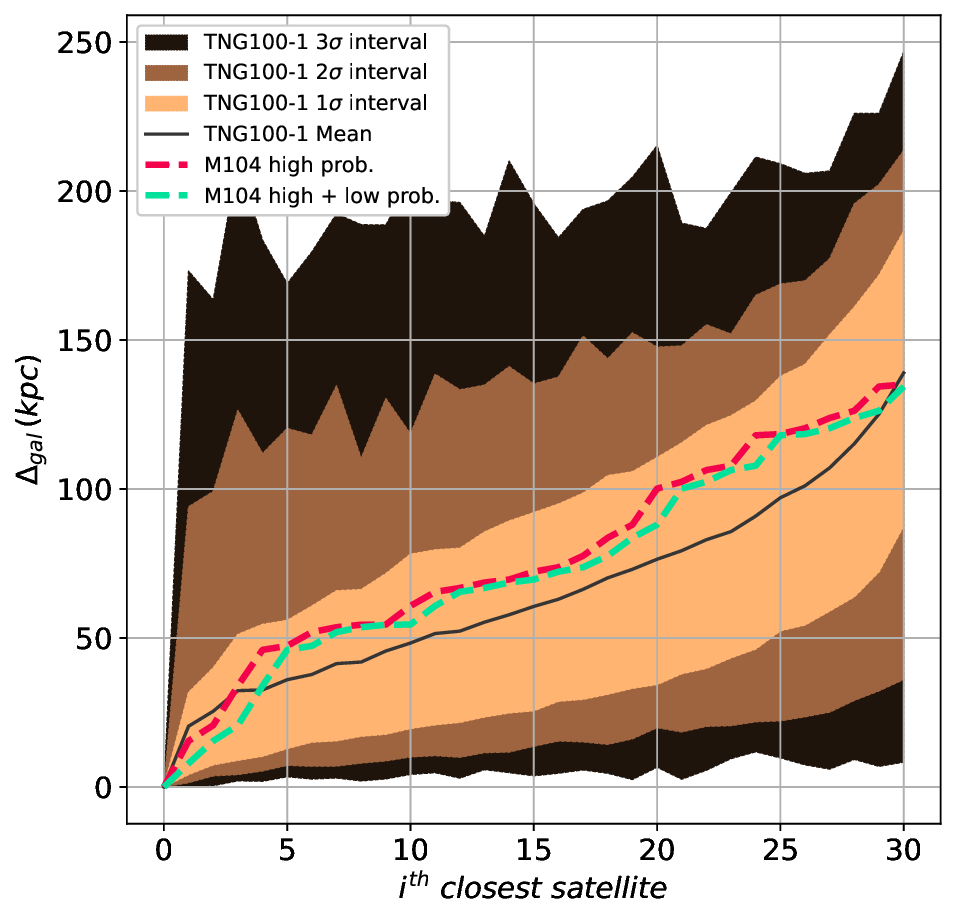}
	\caption{The projected radii ($\Delta_{gal}\,$(kpc)) of the $i^{th}$ closest satellites in the comparison TNG100-1 environments identified in section \ref{M104_sat_plane} as shown by the shaded areas and for M104 itself with the red dashed line (high probability satellites only) and the green dashed line (all satellite candidates).}
	\label{fig:M104_radial_distribution}
\end{figure}

In Figure \ref{fig:M104_radial_distribution}, we show the projected 2D distance of the $i^{th}$ nearest satellite of M104, compared against the distribution of the same simulated environments identified in section \ref{M104_sat_plane}. Overall the radial distribution of satellites around M104 is reproduced well by our selected sample of \textsc{TNG100-1} simulated systems. We fail to find any signal which suggests a reduced density of subhalos at small pericentres. This is expected, the anomaly as discussed lies in satellites radial distance from their hosts, which requires 3D positions of the satellites to properly measure. Here we are forced to consider the 2D projected radii of the satellites instead, therefore any differences in density at small radii between observations and simulations could not be measured reliably. If numerical issues in the simulations are biasing these comparisons, we are unable to quantify this bias with 2D projected sky coordinates alone.

\begin{table*}
\caption{Fundamental parameters and spatial information of all satellites and candidates of the M104 system. Photometric and structural properties are extracted from best-fitting GALFIT models as described in section \ref{photometric_modelling}. Each column is as follows, (1) Galaxy Name, 
(2) Right Ascension (J2000),
(3) Declination (J2000), 
(4) Galaxy Morphology, 
(5) total \textit{g}-band apparent magnitude, 
(6) \textit{g}-band absolute magnitude corrected for Galactic extinction, 
(7) half-light radius in arcsec, 
(8) half-light radius in kiloparsec, 
(9) mean \textit{g}-band surface brightness within the effective radius, 
(10) axis ratio $b/a$,
(11) S\'ersic index,  
(12) projected  distance from M104, 
(13) group membership probability,
(14) low membership probability justification.
For all satellites we assume the distance modulus is that of M104, $\mu = 29.9\pm0.03$ mag \citep{McQuinn_2016}. 
}
\medskip
\resizebox{\textwidth}{!}{%
\tabcolsep=0.0pt
\begin{tabular}{*{15}{c@{\hspace{1em}}}c} 
	\hline
    Galaxy & R.A. & DEC & Morph & $m_g$ & $M_{g,0}$ & $r_e$ & $r_e$ & $\langle\mu{_e,g}\rangle$ & ax. ratio & $n$ & $\Delta_{M104}$ & Memb. & Just. \\
    
    name & hh:mm:ss & hh:mm:ss & Type & mag & mag & arcsec & kpc & mag & - & - & kpc  & Prob. & - \\
    
    (1) & (2) & (3) & (4) & (5) & (6) & (7) & (8) & (9) & (10) & (11) & (12) & (13) & (14)\\
    
    \hline

   UGCA287 & 12:33:55 & -10:40:48 & dIrr & $14.39$ & $-15.63$ & $30.64$ & $1.42$ & $23.82$ & 0.53 & 0.85 & 297.92 & conf & - \\

dw1234-1238 & 12:34:49 & -12:38:24 & dSph & $18.98$ & $-11.15$ & $12.39$ & $0.57$ & $26.44$ & 0.71 & 0.68 & 274.49 & high & - \\

dw1234-1142 & 12:34:49 & -11:42:25 & dTran & $17.3$ & $-12.77$ & $12.25$ & $0.57$ & $24.74$ & 0.65 & 0.74 & 216.34 & high & - \\

dw1235-1247 & 12:35:11 & -12:47:42 & dSph & $21.45$ & $-8.65$ & $6.56$ & $0.3$ & $27.53$ & 0.46 & 0.46 & 280.1 & low & iii. \\

LV J1235-1104 & 12:35:39 & -11:04:01 & BCD & $15.31$ & $-14.72$ & $5.99$ & $0.28$ & $21.19$ & 0.7 & 0.55 & 203.08 & conf & - \\

dw1235-1216 & 12:35:43 & -12:16:23 & dIrr & $17.86$ & $-12.26$ & $10.79$ & $0.5$ & $25.02$ & 0.4 & 0.4 & 208.55 & high & - \\

dw1235-1155 & 12:35:52 & -11:55:52 & dSph & $17.65$ & $-12.45$ & $9.17$ & $0.42$ & $24.46$ & 0.57 & 0.61 & 179.69 & high & - \\

dw1237-1006 & 12:37:10 & -10:06:58 & dSph & $18.03$ & $-11.97$ & $15.69$ & $0.73$ & $26.0$ & 0.46 & 0.77 & 277.37 & high & - \\

dw1237-1125 & 12:37:12 & -11:25:59 & dSph & $18.33$ & $-11.71$ & $8.0$ & $0.37$ & $24.84$ & 0.7 & 0.78 & 120.74 & high & - \\

PGC 042120 & 12:37:14 & -10:29:46 & dIrr & $15.66$ & $-14.34$ & $32.9$ & $1.52$ & $25.24$ & 0.44 & 0.73 & 220.37 & conf & - \\

dw1237-1110 & 12:37:42 & -11:10:08 & dIrr & $18.44$ & $-11.32$ & $6.71$ & $0.31$ & $24.83$ & 0.51 & 1.25 & 121.86 & low & i., iii., v. \\

dw1238-1208 & 12:38:22 & -12:08:06 & dSph & $22.77$ & $-7.27$ & $3.06$ & $0.14$ & $27.19$ & 0.81 & 0.76 & 108.81 & low & ii., iii. \\

dw1238-1120 & 12:38:30 & -11:20:10 & dSph & $22.07$ & $-7.96$ & $3.24$ & $0.15$ & $26.62$ & 0.7 & 0.52 & 78.58 & low & ii., iii. \\

dw1238-1116 & 12:38:31 & -11:16:26 & dSph & $21.31$ & $-8.72$ & $5.81$ & $0.27$ & $27.12$ & 0.78 & 0.7 & 84.54 & low & iii. \\

KKSG31 & 12:38:33 & -10:29:24 & N-dSph & $16.73$ & $-13.56$ & $24.26$ & $1.12$ & $25.36$ & 0.9 & 0.9 & 198.12 & high & - \\

dw1238-1122 & 12:38:34 & -11:22:05 & dIrr & $19.2$ & $-10.83$ & $5.78$ & $0.27$ & $25.01$ & 0.64 & 0.84 & 73.26 & high & - \\

dw1238-1043 & 12:38:42 & -10:43:30 & dSph & $21.67$ & $-8.35$ & $5.37$ & $0.25$ & $27.31$ & 0.52 & 0.59 & 159.19 & high & - \\

dw1238-1102 & 12:38:58 & -11:02:10 & dSph & $20.88$ & $-9.14$ & $5.08$ & $0.24$ & $26.4$ & 0.79 & 0.55 & 106.68 & low & i., iii. \\

dw1239-1227 & 12:39:09 & -12:27:14 & dSph & $21.56$ & $-8.54$ & $4.48$ & $0.21$ & $26.81$ & 0.96 & 0.57 & 142.91 & high & - \\

dw1239-1152 & 12:39:09 & -11:52:37 & dSph & $21.7$ & $-8.36$ & $7.37$ & $0.34$ & $28.03$ & 0.47 & 0.25 & 55.04 & low & iii. \\

dw1239-1159 & 12:39:09 & -11:59:13 & dSph & $19.24$ & $-10.81$ & $13.34$ & $0.62$ & $26.87$ & 0.88 & 1.14 & 70.04 & high & - \\

dw1239-1143 & 12:39:15 & -11:43:08 & N-dSph & $16.45$ & $-13.7$ & $15.82$ & $0.73$ & $24.36$ & 0.66 & 0.91 & 34.59 & high & - \\

dw1239-1154 & 12:39:22 & -11:54:25 & dSph & $21.42$ & $-9.22$ & $7.05$ & $0.33$ & $27.08$ & 0.9 & 0.4 & 53.93 & high & - \\

dw1239-1240 & 12:39:30 & -12:40:30 & N-dSph & $16.48$ & $-13.6$ & $17.79$ & $0.82$ & $24.71$ & 0.49 & 1.17 & 176.56 & high & - \\

NGC4594-DGSAT-3 & 12:39:33 & -11:13:34 & dSph & $18.08$ & $-11.95$ & $16.42$ & $0.76$ & $26.15$ & 0.91 & 0.6 & 68.71 & high & - \\

dw1239-1118 & 12:39:37 & -11:18:32 & dSph & $22.12$ & $-7.91$ & $3.81$ & $0.18$ & $27.02$ & 0.68 & 0.46 & 54.54 & low & ii., iii. \\

dw1239-1106 & 12:39:42 & -11:05:60 & dSph & $21.15$ & $-8.88$ & $4.64$ & $0.21$ & $26.48$ & 0.64 & 0.51 & 88.04 & low & iii. \\

dw1239-1012 & 12:39:43 & -10:12:07 & dSph & $18.23$ & $-11.8$ & $5.68$ & $0.26$ & $24.0$ & 0.94 & 1.44 & 237.17 & low & i., iii. \\

dw1239-1026 & 12:39:51 & -10:26:17 & dIrr & $20.17$ & $-9.85$ & $7.88$ & $0.36$ & $26.65$ & 0.61 & 0.99 & 197.63 & high & - \\

NGC4594-DGSAT-2 & 12:39:51 & -11:20:28 & dSph & $19.61$ & $-10.42$ & $6.9$ & $0.32$ & $25.8$ & 0.89 & 0.73 & 47.34 & high & - \\

NGC4594-DGSAT-1 & 12:39:55 & -11:44:46 & dSph & $17.15$ & $-12.91$ & $28.36$ & $1.31$ & $26.41$ & 0.75 & 1.03 & 20.74 & high & - \\

SUCD1 & 12:40:03 & -11:40:05 & UCD & $18.06$ & $-12.01$ & $1.13$ & $0.05$ & $20.32$ & 0.93 & 0.86 & 7.91 & conf & - \\

dw1240-1323 & 12:40:08 & -13:23:56 & dSph & $19.75$ & $-10.31$ & $3.32$ & $0.15$ & $24.35$ & 0.79 & 0.82 & 296.17 & low & iii., v. \\

KKSG33 & 12:40:09 & -12:21:54 & dSph & $17.79$ & $-12.27$ & $14.2$ & $0.66$ & $25.55$ & 0.92 & 0.61 & 123.84 & high & - \\

dw1240-1118 & 12:40:09 & -11:18:50 & N-dSph & $15.87$ & $-14.19$ & $15.53$ & $0.72$ & $23.8$ & 0.94 & 1.17 & 51.96 & high & - \\

dw1240-1038 & 12:40:14 & -10:38:46 & dIrr & $20.5$ & $-9.47$ & $8.24$ & $0.38$ & $27.13$ & 0.52 & 1.07 & 163.16 & low & i., iii., v. \\

dw1240-1140 & 12:40:18 & -11:40:44 & dSph & $20.46$ & $-9.61$ & $7.42$ & $0.34$ & $26.81$ & 0.87 & 0.28 & 15.6 & low & vi. \\

dw1240-1010 & 12:40:32 & -10:10:23 & dIrr & $19.75$ & $-10.28$ & $8.23$ & $0.38$ & $26.32$ & 0.7 & 0.87 & 242.81 & low & iii., v., vi. \\

dw1240-1024 & 12:40:39 & -10:24:11 & dSph & $20.1$ & $-9.92$ & $5.71$ & $0.26$ & $25.88$ & 0.53 & 0.75 & 205.19 & low & i., iii. \\

dw1240-1245 & 12:40:48 & -12:45:47 & dSph & $20.15$ & $-9.9$ & $3.47$ & $0.16$ & $24.85$ & 0.97 & 0.86 & 192.94 & low & iii., v. \\

dw1240-1012 & 12:40:48 & -10:12:14 & dSph & $20.91$ & $-9.13$ & $3.59$ & $0.17$ & $25.68$ & 0.82 & 0.98 & 238.93 & low & iii. \\

dw1240-1155 & 12:40:60 & -11:55:48 & dIrr & $19.0$ & $-11.06$ & $4.14$ & $0.19$ & $24.08$ & 0.64 & 0.62 & 65.99 & low & iii., v. \\

dw1241-1131 & 12:41:03 & -11:31:41 & N-dSph & $19.1$ & $-10.97$ & $12.74$ & $0.59$ & $26.6$ & 0.96 & 0.7 & 46.77 & high & - \\

dw1241-1210 & 12:41:03 & -12:10:48 & dSph & $19.82$ & $-10.26$ & $3.73$ & $0.17$ & $24.67$ & 0.61 & 0.72 & 102.81 & low & i., iii., v. \\

dw1241-1123 & 12:41:10 & -11:23:53 & dSph & $21.02$ & $-9.04$ & $8.11$ & $0.38$ & $27.56$ & 0.95 & 1.21 & 61.44 & low & iii. \\

	\hline
\end{tabular}}
\label{tab:photometry}
\end{table*}

\begin{table*}
\caption{Table \ref{tab:photometry} continued.}
\medskip
\resizebox{\textwidth}{!}{%
\tabcolsep=0.0pt
\begin{tabular}{*{15}{c@{\hspace{1em}}}c} 
	\hline
    Galaxy & R.A. & DEC & Morph & $m_g$ & $M_{g,0}$ & $r_e$ & $r_e$ & $\langle\mu{_e,g}\rangle$ & ax. ratio & $n$ & $\Delta_{M104}$ & Memb. & Just. \\
    
    name & hh:mm:ss & hh:mm:ss & Type & mag & mag & arcsec & kpc & mag & - & - & kpc  & Prob. & - \\
    
    (1) & (2) & (3) & (4) & (5) & (6) & (7) & (8) & (9) & (10) & (11) & (12) & (13) & (14)\\
    
    \hline

dw1241-1105 & 12:41:10 & -11:05:49 & dSph & $21.83$ & $-8.16$ & $2.69$ & $0.12$ & $26.01$ & 0.81 & 0.71 & 100.52 & low & ii., iii. \\

dw1241-1153 & 12:41:12 & -11:53:31 & dSph & $18.29$ & $-11.78$ & $16.44$ & $0.76$ & $26.37$ & 0.92 & 0.72 & 67.54 & high & - \\

dw1241-1008 & 12:41:17 & -10:08:46 & dSph & $19.9$ & $-10.12$ & $5.83$ & $0.27$ & $25.72$ & 0.83 & 1.06 & 251.99 & high & - \\

KKSG34 & 12:41:19 & -11:55:30 & N-dSph & $17.43$ & $-12.65$ & $17.9$ & $0.83$ & $25.69$ & 0.94 & 0.56 & 74.56 & high & - \\

dw1241-1055 & 12:41:38 & -10:55:34 & dSph & $19.61$ & $-10.42$ & $17.84$ & $0.83$ & $27.86$ & 0.8 & 0.8 & 134.96 & high & - \\

dw1241-1234 & 12:41:48 & -12:34:12 & dSph & $21.16$ & $-8.91$ & $5.11$ & $0.24$ & $26.7$ & 0.78 & 0.5 & 174.77 & high & - \\

dw1242-1116 & 12:42:44 & -11:16:26 & N-dSph & $19.01$ & $-11.07$ & $11.65$ & $0.54$ & $26.33$ & 0.76 & 0.89 & 128.14 & high & - \\

dw1242-1309 & 12:42:45 & -13:09:58 & dSph & $20.81$ & $-9.26$ & $10.96$ & $0.51$ & $28.0$ & 0.54 & 0.91 & 281.65 & high & - \\

PGC42730 & 12:42:49 & -12:23:24 & dSph & $13.72$ & $-16.36$ & $35.44$ & $1.64$ & $23.45$ & 0.71 & 1.28 & 173.77 & conf & - \\

dw1242-1129 & 12:42:50 & -11:29:20 & dSph & $21.18$ & $-8.9$ & $3.63$ & $0.17$ & $25.98$ & 0.78 & 0.46 & 120.27 & low & iii., v. \\

dw1242-1107 & 12:42:56 & -11:07:41 & dSph & $19.21$ & $-10.83$ & $16.52$ & $0.76$ & $27.3$ & 0.91 & 0.86 & 147.57 & high & - \\

dw1242-1010 & 12:42:57 & -10:10:08 & dIrr & $20.97$ & $-9.05$ & $7.98$ & $0.37$ & $27.47$ & 0.69 & 0.53 & 271.83 & high & - \\

dw1243-1137 & 12:43:18 & -11:37:30 & dSph & $21.16$ & $-8.91$ & $7.96$ & $0.37$ & $27.66$ & 0.5 & 1.33 & 137.85 & low & iii. \\

dw1243-1050 & 12:43:33 & -10:50:56 & dSph & $19.01$ & $-11.03$ & $5.27$ & $0.24$ & $24.61$ & 0.57 & 1.29 & 196.73 & low & i., iii. \\

dw1243-1048 & 12:43:38 & -10:48:07 & dSph & $20.95$ & $-9.09$ & $2.28$ & $0.11$ & $24.73$ & 0.91 & 0.66 & 204.55 & low & ii., iii., v. \\

dw1244-1037 & 12:44:13 & -10:37:34 & dIrr & $19.15$ & $-10.91$ & $6.08$ & $0.28$ & $25.06$ & 0.75 & 1.05 & 242.33 & low & iii., v. \\

dw1244-1043 & 12:44:15 & -10:43:19 & dSph & $20.94$ & $-9.12$ & $5.64$ & $0.26$ & $26.69$ & 0.84 & 0.69 & 232.29 & high & - \\

dw1244-1246 & 12:44:17 & -12:46:23 & dSph & $20.39$ & $-9.68$ & $8.32$ & $0.39$ & $26.99$ & 0.82 & 0.52 & 262.08 & high & - \\

dw1244-1135 & 12:44:30 & -11:35:06 & dIrr & $20.85$ & $-9.22$ & $5.72$ & $0.26$ & $26.63$ & 0.45 & 0.65 & 188.14 & high & - \\

dw1244-1138 & 12:44:32 & -11:38:53 & dSph & $22.45$ & $-7.61$ & $2.89$ & $0.13$ & $26.75$ & 0.92 & 0.89 & 189.58 & low & ii., iii. \\

dw1244-1127 & 12:44:38 & -11:27:11 & dIrr & $16.5$ & $-13.56$ & $13.46$ & $0.62$ & $24.14$ & 0.5 & 0.5 & 195.27 & low & i. \\

dw1244-1238 & 12:44:54 & -12:38:10 & dSph & $19.14$ & $-10.92$ & $12.79$ & $0.59$ & $26.67$ & 0.68 & 0.72 & 265.5 & high & - \\

dw1245-1041 & 12:45:07 & -10:41:56 & dTran & $18.44$ & $-11.59$ & $7.38$ & $0.34$ & $24.78$ & 0.59 & 0.9 & 263.44 & high & - \\

dw1246-1104 & 12:46:04 & -11:04:55 & dTran & $16.46$ & $-13.6$ & $4.96$ & $0.23$ & $21.93$ & 0.57 & 0.6 & 268.66 & low & i.,ii., v. \\

dw1246-1240 & 12:46:14 & -12:40:44 & dSph & $19.42$ & $-10.64$ & $7.2$ & $0.33$ & $25.7$ & 0.59 & 0.75 & 314.33 & high & - \\

dw1246-1139 & 12:46:26 & -11:39:00 & dSph & $21.31$ & $-8.73$ & $3.52$ & $0.16$ & $26.04$ & 0.88 & 0.53 & 268.63 & low & iii. \\

dw1246-1108 & 12:46:43 & -11:08:17 & dSph & $21.02$ & $-9.02$ & $4.07$ & $0.19$ & $26.06$ & 0.75 & 0.56 & 291.85 & high & - \\

dw1246-1142 & 12:46:58 & -11:42:25 & dSph & $20.2$ & $-9.85$ & $9.1$ & $0.42$ & $26.99$ & 0.48 & 0.64 & 290.95 & high & - \\

KKSG37 & 12:48:01 & -12:39:18 & N-dSph & $16.87$ & $-13.18$ & $24.9$ & $1.15$ & $25.87$ & 0.77 & 0.77 & 376.44 & high & - \\

dw1248-1037 & 12:48:06 & -10:37:19 & dSph & $21.12$ & $-8.94$ & $7.52$ & $0.35$ & $27.5$ & 0.34 & 0.96 & 376.64 & low & iii. \\

	\hline
\end{tabular}}
\label{tab:photometry2}
\end{table*}

\section{Summary and Conclusion}

We present 40 new satellites galaxy candidate in addition to 35 already known satellite galaxy candidates for the host galaxy M104, where 22 of these new candidates have a high probability of being real satellite galaxies of M104. These candidates are found within our HSC \textit{g}-band images which extend to approximately the virial radius of M104, $420\,$kpc. Detections are complete to a surface brightness limit of $\langle\mu{_e,g}\rangle \approx27.5\,\text{mag}\,\text{arcsec}^{-2}$ or 100 percent complete for objects $M_{\text{g}}<-9$, and 50 percent complete for $-9<M_{\text{g}}<-8$, excluding compact objects of $r_{\text{e}}<300\,$pc.

The new satellite candidates are comparable to other Local Volumes satellites from the ELVES survey \citep{Carlsten_2022}, in terms of structural parameters and photometry. M104 however does possess two unusually compact galaxies, SUCD1 and LV J1235-1104, which we would not have ordinarily recognised as a satellite candidate, due to the difficulty in discriminating them from background galaxies. Therefore, our reported sample would miss similar compact galaxies, if they exist. This isn't a challenge unique to this survey, any dwarf galaxy survey using ground based optical telescopes will have difficulty differentiating compact dwarf galaxies from background galaxies or foreground stars.

We use a simplified PCA approach to measure the 2D flattening of the ellipse best fit to the distribution of the satellites in the sky. We find that with all the satellite candidates of M104, this ellipse is nearly circular, with $b/a=0.97$. This is more circular than all M104 equivalent environments from the Illustris TNG100-1 simulation and even more circular than a purely isotropic distribution of satellites. This could hint at a satellite plane viewed face on. However, if we remove the low probability candidates, this axis ratio drops to $b/a=0.68$, which is then consistent with both the TNG100-1 simulation and an isotropic distribution. This indicates that the low probability candidates may not fill the same distribution of satellites as the high probability candidates, and that they are not bound to M104. The observed circularity of the distribution of all satellites is likely a result of background contamination. Figure \ref{fig:M104_coverage} appears to confirm this, where a number of our newly reported low probability candidates are clustered around a few background galaxies which are all located on one side of the field around M104. We also measured the lopsidedness in the distribution of these satellites using the length of the average lopside vector and found that similarly to the axis ratio, when all satellites are included M104 is lopsided with as a 2$\sigma$ outlier from TNG100-1 and isotropic simulations, but that when only the high probabiliy candidates are included, is consistent with both the TNG100-1 and isotropic simulations. The difference in conclusions based on the inclusion or exclusion of low probability satellite candidates highlights the need for further spectroscopic follow-up, to confirm these candidates and to begin to explore the 3D distribution of these satellites.

Finally, we compared the cumulative satellite luminosity function (CSLF) for M104's satellites within the virial radius to other Local Volume galaxies and found it to closely match the profile of Centaurus A, where both galaxies have no satellites brighter than $M_g<-16.4$ ($L_{g}\sim3*10^{8}\,L_\odot$). Other Local Volume hosts often possess satellites as brighter as this, and sometimes as bright as $M_g=-20$. We performed a comparison of M104 to similar galaxy environments within the TNG100-1 simulation and found M104 is a $2.3\,\sigma$ outlier with the absence of luminous satellites with $M_g<-14$, but the CSLF rises quickly to within expectations at $M_g>-14$. This could be related to the too-big-to-fail problem which describes the absence of high mass satellites when comparing observations to $\Lambda$CDM simulations.

\section*{Acknowledgements}
E.C. and H.J. acknowledge financial support from the Australian Research Council through the Discovery Project DP150100862.

O.M. is grateful to the Swiss National Science Foundation for financial support under the grant number PZ00P2\_202104.

M.S.P. acknowledges funding via a Leibniz-Junior Research Group (project number J94/2020)

This research has made use of the NASA/IPAC Extragalactic Database, which is funded by the National Aeronautics and Space Administration and operated by the California Institute of Technology.

This research has made use of the SIMBAD database, operated at CDS, Strasbourg, France.

We acknowledge the work of Markus Dirnberger to process the optical telescope images used in this paper.


\section*{Data Availability}
The data underlying this article will be shared on reasonable request to the corresponding author.

Additional data underlying this article were derived from sources in the public domain and are available at: https://ned.ipac.caltech.edu/ and https://simbad.u-strasbg.fr/simbad/



\bibliographystyle{mnras}
\bibliography{M104.bib} 


\clearpage 
\appendix

\bsp	
\label{lastpage}
\end{document}